\def\be{\begin{equation}}
\def\ee{\end{equation}}
\def\ba{\begin{eqnarray}}
\def\ea{\end{eqnarray}}
\def\nn{\nonumber}

\def\ellmax{\ell_{\rm max}}

\def\n{{\widehat{\bf n}}}
\def\bigoh{{\mathcal O}}
\def\htau{{\widehat\tau}}
\def\hphi{{\widehat\phi}}
\def\Var{\mbox{Var}}

\def\k{{\mathbf k}}
\def\hk{{\widehat{\mathbf k}}}
\def\n{{\widehat{\mathbf n}}}

\def\hC{{\widehat C}}
\def\hN{{\widehat N}}

\def\hE{\widehat{\mathcal E}}

\def\imp{{\rm impr}}
\def\fsky{f_{\rm sky}}
\def\Spol{S_{\rm pol}}
\def\zmin{z_{\rm min}}
\def\zmax{z_{\rm max}}
\def\chimin{\chi_{\rm min}}
\def\chimax{\chi_{\rm max}}

\documentclass[twocolumn,amsmath,amssymb,prd,floats,aps,psfig]{revtex4}
\usepackage{graphicx, epsfig, natbib}
\usepackage{subfigure}  
\usepackage{color}
\usepackage{epsf}

\newcommand{\wj}[6]{\left(
                           \begin{array}{ccc}
        \! #1\! & #2\!  & #3\!  \\
        \! #4\! & #5\!  & #6\!
                           \end{array}
                   \right)}

\begin{document}

\title{Reconstructing Patchy Reionization from the Cosmic Microwave Background}
	\author{}
\author{Cora Dvorkin$^{1,2}$ and Kendrick M. Smith$^3$}

\affiliation{{}$^1$Kavli Institute for Cosmological Physics and Department of Physics,
University of Chicago, Chicago IL 60637, U.S.A. \\
{}$^2$Enrico Fermi Institute, University of Chicago, Chicago IL 60637, U.S.A.\\
$^3$ Institute of Astronomy, University of Cambridge, CB3 0HA UK}

\baselineskip 11pt
\date{\today}
\begin{abstract}
We introduce a new statistical technique for extracting the inhomogeneous reionization signal from future high-sensitivity
measurements of the cosmic microwave background temperature and polarization fields.  If reionization is inhomogeneous, then the optical
depth to recombination will be a function $\tau(\n)$ of position on the sky.  Anisotropies in $\tau(\n)$ alter the statistics of
the observed CMB via several physical mechanisms: screening of the surface of last scattering, generation of new polarization via
Thomson scattering from reionization bubbles, and the kinetic Sunyaev-Zel'dovich effect.
We construct a quadratic estimator  $\htau_{\ell m}$ for the modes of the $\tau$-field.
This estimator separates the patchy
reionization signal from the CMB in the form of a noisy map, which can be cross-correlated with other probes of reionization
or used as a standalone probe.
A future satellite experiment with sufficient sensitivity and resolution to measure the lensed B-modes on most of the sky can 
constrain key parameters of patchy reionization, such as the duration of the patchy epoch or the mean bubble radius, at
the $\sim 10$\% level.
\end{abstract}

\maketitle

\section{Introduction}

Upcoming generations of cosmic microwave background (CMB) experiments will make precise measurements of secondary anisotropies on small scales $(2000\lesssim\ell\lesssim 10000)$ in temperature and E-mode polarization and will also have sufficient sensitivity to measure secondary B-modes, for example the ones arising from gravitational lensing of the primary E-mode.

Secondary anisotropies are caused by fluctuations in the distribution of matter after recombination and are generated by gravitational lensing, inverse Compton scattering of photons by the hot intracluster medium (the thermal Sunyaev-Zel'dovich effect), and the Doppler effect produced by Thomson scattering of photons from radially moving electrons (the kinetic Sunyaev-Zel'dovich effect (kSZ)). 

Secondary anisotropies are also generated during the inhomogeneous, or patchy, phase of reionization.
Analytic studies and numerical simulations suggest that the period of reionization was more complex than a sudden change in the ionization fraction \cite{Zahn:2006sg,Furlanetto:2004nt,Barkana:2000fd}. 
Reionization is an inhomogeneous process and the inhomogeneities contribute  to the small scale CMB temperature and polarization anisotropies.
In contrast to other secondaries, such as gravitational lensing \cite{Hu:2001fa,Hu:2001kj,Zaldarriaga:1998ar,Okamoto:2003zw,Hirata:2002jy,Hirata:2003ka,Smith:2007rg,Hirata:2008cb} or the Sunyaev-Zel'dovich effect (SZ) \cite{Sunyaev:1970er,Sunyaev:1980vz}, the potential science returns from patchy reionization have not been extensively studied in the context of the CMB.

There are currently almost no observational constraints on the evolution of the ionization fraction during the epoch of reionization.
The Gunn-Peterson trough in high-redshift quasar spectra has been observed, showing that the transition from partial reionization to a fully ionized universe ($\overline x_e\approx 1$) took
place at redshift $z\sim 6$ \cite{Becker:2001ee,White:2003sc,Fan:2005es}. 
Large-scale E-mode polarization measurements from five-year WMAP data show a total optical depth to
recombination $\tau=0.087\pm 0.017$.
If reionization is assumed instantaneous, this would imply a transition redshift $z_{\rm rei}=11.0\pm 1.4$ with $68 \%$ confidence \cite{Dunkley:2008ie}.
Combining these two observations, we therefore have good indirect evidence for patchy reionization but no detailed information.

Predictions from simulations tell us that in order to allow the contributions of the first stars \cite{Ricotti:2004xf}, the beginning of reionization could go until $z\sim 30$.
However, because of a lack of experimental data, this prediction is still uncertain.
The large-scale E-mode is sensitive to reionization history, but only contains information about the (spatial) average ionization fraction $\overline x_e(z)$ during reionization, not the size or morphology of the ionized regions.
The WMAP data are not sensitive enough to place strong constraints on the reionization history beyond determination of the total optical depth $\tau$,
but future large-scale E-mode measurements from Planck or CMBpol can constrain 
up to $\sim 5$ principal components or redshift bins in $\overline x_e(z)$ \cite{Mortonson:2007hq,Hu:2003gh}.

The patchy reionization signal is small and, therefore, will be difficult to separate from the other secondaries using the power spectrum alone. 
The patchy contribution to the temperature power spectrum is smaller than the sum of the contributions from lensing and low-redshift kSZ
on all angular scales \cite{Gnedin:2000gr,Zahn:2005fn,Hu:1999vq,Iliev:2006un,McQuinn:2005ce,Santos:2003jb}. 
The thermal SZ signal from galaxy clusters leads to a larger signal than the one caused by the kSZ effect, 
but it can be separated from the other secondaries due to its non-blackbody frequency dependence.
In polarization, currently favored models produce B-mode polarization power spectra
with an amplitude ($\sim 0.01$$\mu$K)  that is significantly lower than the B-modes coming from gravitational lensing 
($\sim 0.1$$\mu$K) \cite{Gruzinov:1998un,Liu:2001xe,Mortonson:2006re,Dore:2007bz}.

In this paper, we will propose an estimator which isolates the patchy reionization signal in the CMB.
If reionization is patchy, then the optical depth to last scattering will be a 2D field $\tau_{\ell m}$ rather than a constant $\tau$.
Our estimator $\htau_{\ell m}$ will reconstruct each mode of this field, within statistical noise, using the small change in the CMB
statistics which is induced by the mode.
This construction was inspired by the well-known lens reconstruction estimator \cite{Hu:2001fa,Hu:2001kj,Okamoto:2003zw,Hirata:2002jy,Hirata:2003ka},
which reconstructs each mode $\phi_{\ell m}$ of the CMB lens potential in an analogous way.
We will show that our new estimator $\htau_{\ell m}$ isolates the patchy signal, in the sense that its expectation value is simply the
underlying field $\tau_{\ell m}$, with no contribution from the Gaussian part of the CMB.

In \S\ref{sec:reionization_model} and ~\S\ref{sec:reionization_and_the_CMB},
we describe our semi-analytic modeling of reionization and its effect on the CMB.
We will split the signal from patchy reionization into three effects: direction-dependent screening of the acoustic peaks,
polarization generated by Thomson scattering from reionization bubbles, and temperature anisotropy generated by radial motion
of the bubbles (i.e. the kSZ effect).

In \S\ref{sec:simple_estimator} and \S\ref{sec:quad_est_realistic} we will construct the estimator $\htau_{\ell m}$,
by analogy with the quadratic estimator for lens reconstruction.
We will show that the first two effects from \S\ref{sec:reionization_and_the_CMB} (screening and Thomson scattering) 
have an algebraic form where the quadratic estimator
construction applies, but the third effect (kSZ) does not, because the kSZ anisotropy is proportional to the velocity field during
reionization, and this field is not highly correlated with the primary CMB.
Consequently, our estimator $\htau_{\ell m}$ is not very sensitive to the kSZ effect, but should be a near-optimal statistic for
extracting the screening and Thomson signals.
Although a complete treatment which formally extracts all the signal-to-noise is quite complex (\S\ref{ssec:pca}) we show that a simple
estimator (\S\ref{ssec:simple_estimator}) contains all the information in practice.

In \S\ref{sec:simulation} we demonstrate our estimator with Monte Carlo simulations. 
We make simplifying assumptions, such as simulating the lensed component of the $B$-mode as if it were a Gaussian field.
In \S\ref{sec:forecasts} we present our forecasts. 
A future all-sky $B$-mode experiment can constrain 
key reionization parameters, such as the mean radius of the ionized bubbles or the duration of reionization, at the $\sim 10\%$ level.
It might be possible to improve this measurement making use of a delensing procedure or cross-correlating the $\tau$-map with large-scale structure.
We conclude in \S\ref{sec:discussion}.

\section{Reionization model}
\label{sec:reionization_model}

Throughout this paper, we use a fiducial cosmology defined by the WMAP5+BAO+SN parameters from \cite{Komatsu:2008hk}:
\ba
&& \{ \Omega_b h^2, \Omega_m h^2, h, \Omega_\Lambda, \tau, A, n_s \} = \\
&& \{ 
0.02265, 
0.137, 
0.701, 
0.721,
0.084,
2.16 \times 10^{-9},
0.96
\} \nn
\ea

For consistency with CAMB \cite{Lewis:1999bs}, we model the ionization fraction in the following way:
\be\label{eq:xe_form}
\overline x_e(z) = {1 \over 2}\left[1 - \tanh\left({y(z)-y_{re} \over \Delta_y}\right)\right],
\ee
where $y(z)=(1+z)^{3/2}$. $y_{re}$ and $\Delta_y$ are free parameters of the model.
We parameterize the reionization model by taking $\Delta_y$ and the total optical depth $\tau$
to be the model parameters, and treat $y_{re}$ as a derived parameter.

We represent the H II regions by ionized spherical bubbles of radius $R$ with a log-normal distribution \cite{Furlanetto:2004nh,Zahn:2006sg}, with a characteristic size given by $\bar{R}$ and width of the distribution given by $\sigma_{\ln R}$:
\be
P(R) = {1 \over R}{1 \over \sqrt{2\pi\sigma_{\ln R}^2}}e^{{-[\ln({R/\bar{R}})]^2/(2\sigma_{\ln R}^2)}}  \label{eq:log-normal}
\ee
We assume that the number density of bubbles fluctuates as a biased tracer of the large-scale structure with bubble bias $b$.
For simplicity, we will assume that the bias is independent of the bubble radius.

\begin{figure}[htbp]
\begin{center}
\includegraphics[width=4.in]{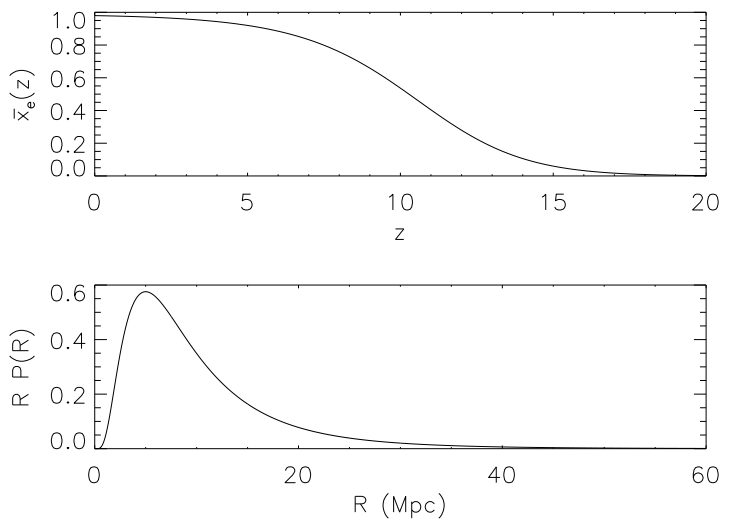}
\end{center}
\caption{Average ionization fraction $\bar x_e(z)$ (upper panel) and the bubble radius distribution $P(R)$ (lower panel) in our fiducial reionization model.}
\label{fig:fiducial_parameters}
\end{figure}

We take our fiducial model to be described by the following set of parameters:
\be
\{\tau, \Delta_y, b, \bar{R}, \sigma_{\ln R}\}
=
\{ 0.084, 19.0, 6.0, 5\mbox{ Mpc}, \ln(2) \}
\ee
In Fig. \ref{fig:fiducial_parameters} we show the evolution of the ionization fraction in redshift (upper panel) and the radius distribution (lower panel) for our fiducial reionization model.

The line-of-sight integral for the optical depth during the inhomogeneous period can be written as 
\be
\tau(\n,z) = \sigma_T n_{p,0} \int_0^z \frac{dz'(1+z')^2}{H(z')} x_e(\n,z'),
\ee
where $H(z)$ is the Hubble parameter, $\sigma_T$ is the Thomson scattering cross section, $n_{p,0}$ is the present number density of protons,
and $x_e(\n,z)$ is the ionization fraction in direction $\n$ at redshift $z$.

Under the Limber approximation, the angular power spectrum of $\tau(\n)$ during the inhomogeneous period of reionization can be written as
\be
C_\ell^{\tau\tau} = \int d\chi\, \frac{\sigma_T^2 n_{p,0}^2}{a^4\chi^2} P_{\Delta x_e\Delta x_e}(\chi,k=\ell/\chi),  \label{eq:cltau}
\ee
where $\Delta x_e$ is the ionization fraction fluctuation and $\chi$ is the comoving distance from the observer along the line-of-sight.

The three-dimensional power spectrum $P_{\Delta x_e\Delta x_e}$ is a sum of two terms \cite{Gruzinov:1998un,Wang:2005my,Mortonson:2006re}, corresponding to $1$-bubble ($1b$) and $2$-bubble contributions ($2b$) to the power spectrum of the ionized hydrogen:
\be\label{eq:pxexe}
P_{\Delta x_e\Delta x_e}(k) = P_{\Delta x_e\Delta x_e}^{{\rm 1b}}(k) + P_{\Delta x_e\Delta x_e}^{{\rm 2b}}(k)
\ee
The $1$-bubble contribution to the three-dimensional power spectrum is given by \cite{Mortonson:2006re}
\be\label{eq:1-bubble}
P_{\Delta x_e\Delta x_e}^{{\rm 1b}}(k)=x_e(1 - x_e) [ F(k)  + G(k) ]\,,
\ee
with the functions $F(k)$ and $G(k)$ defined by
\ba
F(k) &=& \frac{\int dR P(R) [V(R)]^2 [W(kR)]^2}{\int dR P(R) V(R)}\\
G(k) &=& \int \frac{d^3\k'}{(2\pi)^3} P(|\k-\k'|) F(k')
\ea
where $V(R)=4\pi R^3/3$ is the volume of the bubble, $P(R)$ is the log-normal distribution of bubbles defined in Eq. (\ref{eq:log-normal}), 
$P(k)$ is the matter power spectrum, and $W(kR)$ is the Fourier transform of a real-space tophat window function with radius $R$, given by:
\be
W(kR)={3\over(kR)^3}\left[\sin(kR)-kR\cos(kR)\right]
\ee
The first term in Eq. (\ref{eq:1-bubble}) is associated with the shot noise of the bubbles.
The $2$-bubble contribution is given by 
\be\label{eq:2-bubble}
P_{\Delta x_e\Delta x_e}^{{\rm 2b}}(k)=\left[(1 - x_e)\ln(1 - x_e) I(k) - x_e \right]^2 P(k),
\ee
where $I(k)$ is defined by
\be
I(k) = b\, \frac{\int dRP(R)V(R)W(kR)}{\int dRP(R)V(R)},
\ee
and $b$ is the bubble bias.

\begin{figure}[htbp]
\begin{center}
\includegraphics[width=3.5in]{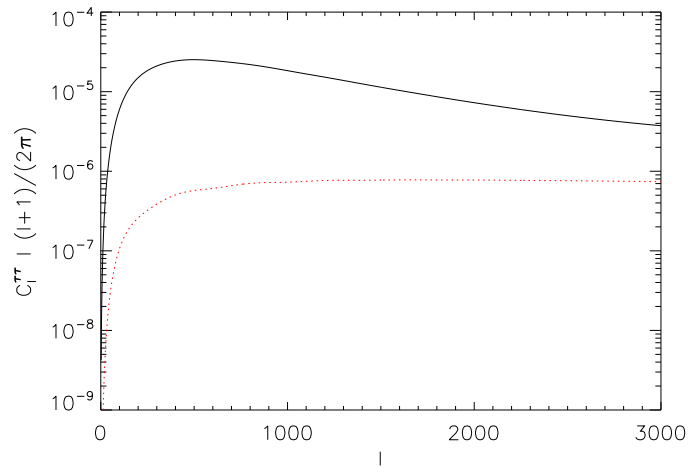}
\end{center}
\caption{Power spectrum $C_\ell^{\tau\tau}$ of the optical depth $\tau(\n)$ (solid/black curve) 
compared to the $x_e^2P(k)$ contribution (see Eq.~(\ref{eq:2-bubble})) to $C_\ell^{\tau\tau}$ alone (dotted/red curve), in our fiducial reionization model.
The $x_e^2P(k)$ term does not vanish as $x_e\rightarrow 1$, and should not be thought of as part of the patchy reionization signal, but it is a small contribution to the power spectrum.}
\label{fig:xe2}
\end{figure}

Note that the 2-bubble contribution does not go to zero as $x_e\rightarrow 1$, due to the $x_e^2 P(k)$ term in Eq. (\ref{eq:2-bubble}).
This term corresponds to fluctuations in the free electron density which are due to matter fluctuations alone, independent of the bubble distribution.
Since it comes from both the homogeneous and inhomogeneous periods of reionization, we will not consider the $x_e^2P(k)$ term to be part of the patchy signal in this paper,
although we also find that this term is a small contribution to the total power spectrum (Fig.~\ref{fig:xe2}).

\section{Reionization and the CMB}
\label{sec:reionization_and_the_CMB}

Inhomogeneous reionization produces three effects in the CMB:
\begin{enumerate}
\item Screening: temperature anisotropy and polarization from the surface of last scattering are multiplied
by $e^{-\tau(\n)}$.  If reionization is inhomogeneous, the screening effect generates B-modes in polarization
because the optical depth $\tau(\n)$ to recombination is a function of the line-of-sight direction $\n$.
\item Thomson scattering: new polarization is generated during the patchy epoch via scattering of the local
CMB {\em temperature} quadrupole by ionized bubbles.  This effect also generates B-modes
\cite{Gruzinov:1998un,Liu:2001xe,Mortonson:2006re,Dore:2007bz}.
\item kSZ: new temperature anisotropy is generated from the radial motion of reionization bubbles relative to
the observer
\cite{Zahn:2005fn,Hu:1999vq,Iliev:2006un,McQuinn:2005ce,Santos:2003jb,Jaffe:1997ye}.
\end{enumerate}
In this section we will write unified expressions (Eqs.~(\ref{eq:sec3_qu}),~(\ref{eq:sec3_t}) below) for the patchy signal
in temperature and polarization.
The physics of each of these three effects will be encoded in ``response fields'' $T_1$ and $E_1$, which will be studied in more detail in subsequent sections.

First, we introduce some notation.
Throughout this section we will parameterize line-of-sight integrals by $\chi$, the comoving distance from the observer.
We denote the total optical depth to distance $\chi$ along the line-of-sight $\n$ by:
\be
\tau(\n,\chi) = \sigma_T n_{p,0} \int_0^\chi \frac{d\chi}{a^2} x_e(\n,\chi)
\ee
To study the patchy reionization signal in polarization, we write the observed polarization $(Q\pm iU)$
as a line-of-sight integral:
\ba
(Q\pm iU)(\n) &=& \int_0^\infty d\chi\, \dot\tau e^{-\tau(\n,\chi)} \Spol^\pm(\n,\chi) \label{eq:line_of_sight_qu} \\
\Spol^\pm(\n,\chi) &=& -\frac{\sqrt{6}}{10} \sum_m ({}_{\pm 2}Y_{2m}(\n)) a_{2m}^{T}(\n,\chi), \nn
\ea
where $a_{2m}^{T}(\n,\chi)$ are the temperature quadrupole moments at each position in space and $({}_{\pm 2}Y_{2m})$ are the spin $\pm2$ spherical harmonics.
Overdots denote derivatives with respect to conformal time.
The integral formally runs from $\chi=0$ to $\chi=\infty$, but
it only receives nonzero contributions from the epochs of reionization ($0\lesssim z\lesssim 20$ in our fiducial model)
and last scattering $(700\lesssim z\lesssim 1300)$.

We write $x_e(\n,\chi)$ as its angular average plus a fluctuation term:
\be
x_e(\n,\chi) = \bar x_e(\chi) + \Delta x_e(\n,\chi)
\ee
Now let us expand $(Q\pm iU)(\n)$ in powers of $\Delta x_e$, keeping zeroth and first order terms.
A formal way of doing this is to write:
\ba
(Q\pm iU)(\n) &=& (Q\pm iU)_0(\n) \label{eq:qu1_integral} \\
&& \hskip -0.4in + \sigma_T n_{p,0} \int \frac{d\chi}{a^2} \Delta x_e(\n,\chi) (Q\pm iU)_1(\n,\chi), \nn
\ea
where $(Q\pm iU)_0$ is the polarization from recombination and homogeneous reionization, and the $\chi$-dependent response field $(Q\pm iU)_1$ is defined by:
\ba
&& (Q\pm iU)_1(\n,\chi) = \int_\chi^\infty d\chi' \frac{\delta (Q\pm iU)(\n)}{\delta \tau(\chi')} \label{eq:line_of_sight_qu1} \\
&& \hskip 0.4in = e^{-\tau(\chi)} \Spol^\pm(\n,\chi) - \int_\chi^\infty d\chi' \dot\tau e^{-\tau} \Spol^\pm(\n,\chi'), \nn
\ea
where $\delta/\delta\tau(\chi')$ in the first line denotes the functional derivative.  (Note that the response field is an E-mode and so we will use the
notation $a^{E_1(\chi)}_{\ell m}$ when we write it in harmonic space.)

The two terms in the second line correspond physically to Thomson scattering and screening respectively.
In the Thomson term, an $x_e$ fluctuation generates new polarization which is proportional to the source term $\Spol^{\pm}$ evaluated at the same point along the line-of-sight.
In the screening term, an $x_e$ fluctuation changes the amplitude of all polarization generated at earlier times,
i.e. from earlier in the patchy epoch or from recombination.

In this paper, we will find it convenient to represent Eq.~(\ref{eq:qu1_integral}) in binned form.  We introduce a set of $N$ redshift
bins which cover the patchy epoch.  We denote the endpoints of the $\alpha$-th bin by $[\zmin^\alpha, \zmax^\alpha]$ and
the central redshift by $z^\alpha$.
We denote the values of $\chi$ which correspond to these values of $z$ by $\chimin^\alpha, \chi^\alpha, \chimax^\alpha$.
(Throughout this section, we have found it convenient to take $\chi$ as the time coordinate, but in subsequent
sections it will be more intuitive to use $z$, so we define our binned representation using redshift bins instead of $\chi$ bins.)

We then write:
\be
(Q\pm iU)(\n) = (Q\pm iU)_0(\n) + \sum_\alpha \Delta\tau^\alpha(\n) (Q\pm iU)^\alpha_1(\n),  \label{eq:sec3_qu}
\ee
where:
\be
\Delta\tau^\alpha(\n) = \sigma_T n_{p,0} \int_{\chimin^{\alpha}}^{\chimax^\alpha} \frac{d\chi}{a^2} \Delta x_e(\n,\chi)  
\ee
\ba\label{eq:qu1_binned}
(Q\pm iU)^\alpha_1(\n) &=&
  e^{-\bar\tau(\chi^\alpha)} \Spol^\pm(\n,\chi^\alpha)    \nn \\
&& -\int_{\chi^\alpha}^\infty d\chi'\, \dot\tau e^{-\tau} \Spol^\pm(\n,\chi') 
\ea
We have assumed here that $\Delta\tau^\alpha(\n) \ll 1$.

Eqs.~(\ref{eq:sec3_qu}) and ~(\ref{eq:qu1_binned}) are the representation for the patchy polarization signal that we will
use throughout this paper.
In this form, we think of patchy reionization as being represented by a set of 2D fields
$\Delta\tau^\alpha(\n)$ which correspond to the optical depth anisotropy in redshift bins.
The CMB polarization signal is obtained by multiplying each of these fields by a ``response field''
$(Q\pm iU)_1^\alpha$, which is defined (Eq.~(\ref{eq:qu1_binned})) in a way which does not depend on the bubble realization.
The response field is a pure E-mode; we will use the notation $a^{E_1 (\alpha)}_{\ell m}$ when we write it in harmonic space.

The temperature case is analogous but the source term is more complicated.
We give an outline of the calculation here; the details can be found in App.~\ref{app:power_spectrum_calc}.
We can still write $T(\n)$ as a line-of-sight integral:
\be
T(\n) = \int_0^\infty d\chi\, S_T(\n,\chi,\tau(\n,\chi)),  \label{eq:s_t_def}
\ee
where $S_T$ is a function of local quantities at $(\n,\chi)$ and the local optical depth $\tau(\n,\chi)$.
(The full expression for $S_T$ is given in Eq.~(\ref{eq:app_los_t}).)
Starting from this line-of-sight integral, we obtain a binned expression:
\ba
T(\n) &=& T_0(\n) + \sum_\alpha \Delta\tau^\alpha(\n) T^\alpha_1(\n)  \label{eq:sec3_t}  \\
T_1^\alpha(\n) &=& \int_{\chi^\alpha}^\infty d\chi'\, \frac{\delta S_T}{\delta\tau(\chi')}  \label{eq:sec3_fd}
\ea
In App.~\ref{app:power_spectrum_calc}, we evaluate the functional derivative and compute the explicit form of the response field $T_1$ (see Eq.~(\ref{eq:t1_horrible})).
The main contributions to $T_1$ are a Doppler term in which a $\tau$ fluctuation generates new temperature anisotropy at the same redshift, and a screening term
in which a $\tau$ fluctuation changes the amplitude of all anisotropies generated at earlier redshifts.

For now we simply note that the patchy signal in temperature can be written in a form (Eq.~(\ref{eq:sec3_t})) which
is analogous to polarization (Eq.~(\ref{eq:sec3_qu})): the optical depth $(\Delta\tau^\alpha)$ in each redshift bin multiplies
a response field $T_1$ which is independent of the realization of patchy reionization.
We will use this representation for the temperature signal throughout the paper.

\begin{figure}
\begin{center}
\includegraphics[width=3.5in]{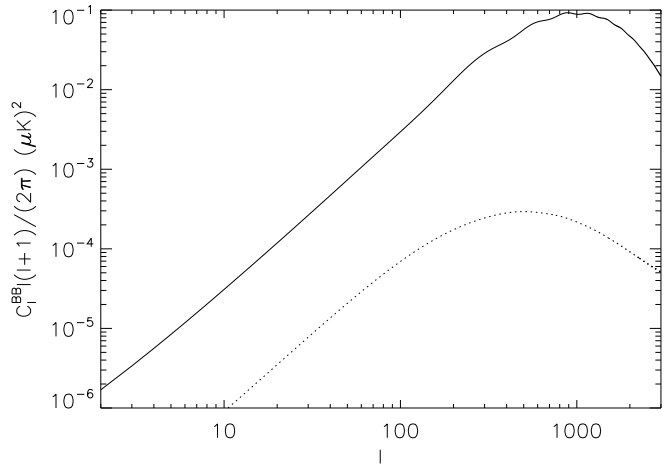}
\end{center}
\caption{Comparison between the B-mode lensing power spectrum (solid curve) and the B-mode power spectrum from patchy reionization (dotted curve) in our fiducial model.}
\label{fig:Bmodes}
\end{figure}

Since we have split the $\Delta\tau$-field into redshift bins, we will need to split the power spectrum $C_\ell^{\tau\tau}$ into bins as well.
In the Limber approximation the power spectrum is diagonal in redshift bins, and given by restricting the integral in Eq.~(\ref{eq:cltau}) 
to the relevant redshift range:
\be
C_\ell^{\tau_\alpha \tau_\beta} = \delta_{\alpha\beta} \sigma_T^2 n_{p,0}^2 \int_{\chimin^\alpha}^{\chimax^\alpha}
\frac{d\chi}{a^4 \chi^2} P_{\Delta x_e\Delta x_e}(\chi,k=\ell/\chi)
\ee
In polarization, inhomogeneous reionization generates a B-mode, in contrast to the homogeneous case.
Under the Limber approximation, valid for $\ell \gg 10$ \cite{Mortonson:2006re,Dore:2007bz}, the B-mode power spectrum 
is related to the power spectrum of the ionized hydrogen, $P_{\Delta x_e\Delta x_e}(k)$ as 
\ba
C_{\ell}^{BB}&=&{3\sigma_T^2 n_{p,0}^2 \over 100} \int_{\chi_{i}}^{\chi_{f}}{d\chi\over a^4\chi^2} \nn\\
&&\times e^{-2\tau(\chi)} Q_{\rm rms}^2(\chi) P_{\Delta x_e\Delta x_e}(\chi,k=\ell/\chi),  \label{eq:clbb}
\ea
where $P_{\Delta x_e\Delta x_e}$ is given by Eqs.~(\ref{eq:pxexe}), (\ref{eq:1-bubble}) and (\ref{eq:2-bubble}), and $\chi_i$, $\chi_f$ are comoving distances to the beginning and end of reionization respectively.
$Q_{\rm rms}(\chi)$ is the rms temperature quadrupole, which we treat as constant during the epoch of patchy reionization, with a 
value of $Q_{\rm rms}=22$$\mu$K in the fiducial model.
In Fig.~\ref{fig:Bmodes} we show the B-mode power spectrum from patchy reionization (Eq.~(\ref{eq:clbb})) with the lensing B-mode shown for comparison.

Throughout this paper, CMB noise power spectra will be given by
\ba
N_{\ell}^{EE}=N_\ell^{BB}=\Delta_P^2 \exp\left( \frac{\ell(\ell+1)\theta_{\rm FWHM}^2}{8\ln(2)} \right)
\ea
where $\Delta_P$ is the detector noise in $\mu$K-steradians and $\theta_{\rm FWHM}$ is the FWHM of the beam in steradians.

\section{Quadratic estimator for patchy reionization: toy example}
\label{sec:simple_estimator}

In the previous subsection, we saw that the CMB anisotropy with the patchy reionization contribution included
can be written in the following form:
\be\label{eq:totalt}
T(\n) = T_0(\n)+\sum_{\alpha}\Delta\tau^{\alpha}(\n)T_1^{\alpha}(\n)   
\ee
\be\label{eq:totalqplusiu}
(Q\pm iU)(\n) = (Q\pm iU)_0(\n)+\sum_{\alpha}\Delta\tau^{\alpha}(\n)(Q\pm iU)_1^{\alpha}(\n) 
\ee
The fields $(\Delta\tau)^\alpha, T_1^\alpha, (Q\pm iU)_1^\alpha$ are not directly observable, but their presence alters the
statistics of the CMB.  The leading terms $T_0, (Q\pm iU)_0$ in Eqs.~(\ref{eq:totalt}),~(\ref{eq:totalqplusiu}) are Gaussian
to a good approximation (ignoring non-Gaussian contributions from CMB lensing which will be discussed later) but the patchy
terms are not Gaussian fields.

How can we construct an estimator which can detect the patchy terms statistically?
In this section and the next, we will propose an answer to this question.
We will construct an estimator $\htau_{\ell m}$ which reconstructs the modes of the $\tau$ field from the CMB.
This estimator will separate the patchy and Gaussian contributions, in the sense that the Gaussian part of the CMB does not
contribute to the expectation value of the estimator (but does act as noise in the reconstruction).

In this section, we will consider a simplified case to build intuition.
We consider polarization only and make the ``constant quadrupole'' approximation: 
we assume that $(Q\pm iU)_1^\alpha = (Q\pm iU)_1^\beta$ for all
pairs of redshift bins $\alpha,\beta$ during the epoch of patchy reionization.
Physically, this approximation means that the source term for CMB polarization,
\be
-\frac{\sqrt{6}}{10} \sum_m ({}_{\pm 2}Y_{2m}(\n)) a_{2m}^{T}(\n,z)
\ee
depends on the line-of-sight direction $\n$ but is independent of the redshift $z$.
In reality, the temperature quadrupole $a_{2m}^{T}(\n,z)$ is not 100\% correlated between
redshifts, but decorrelates on the scale $\Delta z \approx 5$ \cite{Dore:2007bz}.
We will see how to include this decorrelation in our formalism in the next section.

In the constant quadrupole approximation, Eq.~(\ref{eq:totalqplusiu}) becomes:
\be
(Q\pm iU)(\n) = (Q\pm iU)_0(\n) + \Delta\tau(\n)(Q\pm iU)_1(\n),  \label{eq:sec4_onebin}
\ee
where $\Delta\tau(\n) =  \sum_\alpha \Delta\tau^\alpha(\n)$ is the inhomogeneous part of the total optical depth to recombination.
There is a formal analogy between this expression and CMB lensing \cite{Seljak:1995ve,Zaldarriaga:1998ar,Lewis:2006fu}.
The lensed CMB polarization is given by:
\be
(Q\pm iU)_{\rm lensed} = (Q\pm iU)_{\rm unl} + (\nabla\phi)_a \nabla^a(Q\pm iU)_{\rm unl},  \label{eq:sec4_lensing}
\ee
where neither the unlensed CMB $(Q\pm iU)_{\rm unl}$ nor the lens potential $\phi$ are directly observable.
Nevertheless, there is a quadratic estimator $\hphi_{\ell m}$ which ``reconstructs'' the lens potential
mode by mode, using only the observed polarization $(Q\pm iU)_{\rm lensed}$ 
\cite{Hu:2001fa,Hu:2001kj,Okamoto:2003zw,Hirata:2002jy,Hirata:2003ka}.

On a technical level, the lens reconstruction estimator $\hphi_{\ell m}$ is possible because the
two-point function of the lensed CMB contains a term which is linear in $\phi$:
\be
\left\langle a_{\ell_1 m_1}^{E}a_{\ell_2 m_2}^{B}\right\rangle
=
\sum_{\ell m}\Gamma^{EB(\phi)}_{\ell_1\ell_2\ell}\wj{\ell_1}{\ell_2}{\ell}{m_1}{m_2}{m} \phi^{*}_{\ell m}, \label{eq:sec4_eb_lensing}
\ee
where the lensing coupling $\Gamma^{EB(\phi)}_{\ell_1\ell_2\ell}$ is given in App.~\ref{sec:optimal_estimator}
(Eq.~(\ref{eq:gamma_lensing})).
(The expectation value $\langle\cdot\rangle$ in Eq.~(\ref{eq:sec4_eb_lensing}) is taken over realizations of the CMB
in a fixed realization of the lens potential $\phi$.)

Analogously, in the constant quadrupole approximation of this section,
patchy reionization induces a two-point function which is linear in the
field $\Delta\tau$:
\be
\left\langle a_{\ell_1 m_1}^{E}a_{\ell_2 m_2}^{B}\right\rangle=\sum_{\ell m}\Gamma^{EB(\tau)}_{\ell_1\ell_2\ell}\wj{\ell_1}{\ell_2}{\ell}{m_1}{m_2}{m} (\Delta\tau)^{*}_{\ell m},  \label{eq:sec4_twopoint}
\ee
where the patchy coupling $\Gamma^{EB(\tau)}_{\ell_1\ell_2\ell}$ is given by:
\ba
\Gamma^{EB(\tau)}_{\ell_1\ell_2\ell}={C_{\ell_1}^{E_{0}E_{1}}\over 2i}\sqrt{{(2\ell_1+1)(2\ell_2+1)(2\ell+1) \over 4\pi}} \nn\\
\times\left[\wj{\ell_1}{\ell_2}{\ell}{-2}{2}{0}-\wj{\ell_1}{\ell_2}{\ell}{2}{-2}{0}\right]  \label{eq:sec4_gamma_eb}
\ea
Starting from this two-point function, one can construct a minimum-variance quadratic estimator $\htau_{\ell m}$ for
the field $\Delta\tau$:
\ba\label{eq:quadratic_estimator}
\htau^*_{\ell m} &=& N_\ell^{\tau\tau} \sum_{\ell_1\ell_2}^{\ell_{\rm max}}\sum_{m_1m_2}
{\Gamma^{EB(\tau)*}_{\ell_1\ell_2\ell}\over(C_{\ell_1}^{EE}+N_{\ell_1}^{EE})(C_{\ell_2}^{BB}+N_{\ell_2}^{BB})} \nn\\
&&\times \wj{\ell_1}{\ell_2}{\ell}{m_1}{m_2}{m}
a_{\ell_1m_1}^{E} a_{\ell_2m_2}^{B},  
\ea
where
\be
N_\ell^{\tau\tau} = \left[ \frac{1}{2\ell+1} \sum_{\ell_1\ell_2}
\frac{|\Gamma^{EB(\tau)}_{\ell_1\ell_2\ell}|^2}{(C_{\ell_1}^{EE}+N_{\ell_1}^{EE})(C_{\ell_2}^{BB}+N_{\ell_2}^{BB})}
\right]^{-1}  \label{eq:sec4_nl}
\ee
The derivation of this estimator is given in App.~\ref{sec:optimal_estimator}.

The estimator $\htau_{\ell m}$ is unbiased, in the sense that:
\be
\langle \htau_{\ell m} \rangle = (\Delta\tau)_{\ell m},  \label{eq:sec4_unbiased}
\ee
and $N_\ell^{\tau\tau}$ is the noise power spectrum of the reconstruction, 
\be
\langle \htau^*_{\ell m} \htau_{\ell' m'} \rangle_{\rm noise} = N_\ell^{\tau\tau}\delta_{\ell\ell'}\delta_{mm'},  \label{eq:sec4_noise_covariance}
\ee
where $\langle\cdot\rangle_{\rm noise}$ denotes an expectation value taken over CMB realizations containing no patchy signal
(i.e. Gaussian realizations).  The expectation value on the LHS of Eq.~(\ref{eq:sec4_nl}) is what we mean by 
``reconstruction noise'' power in the context of $\tau$ reconstruction; it is the power spectrum of the reconstruction
due to statistical fluctuations alone, in the absence of any patchy signal.

The construction of $\htau_{\ell m}$ in Eq.~(\ref{eq:quadratic_estimator}) 
is formally identical (with a different $\Gamma$ coupling) to the lens
reconstruction estimator $\hphi_{\ell m}$; details are given in App.~\ref{sec:optimal_estimator}.
However, there is one qualitative difference between the patchy quadratic estimator $\htau_{\ell m}$ and
the lensing estimator $\hphi_{\ell m}$ that we would like to emphasize.
The polarization signal from lensing is proportional to the unlensed CMB (Eq.~(\ref{eq:sec4_lensing}))
whereas the patchy signal is proportional to a new field $E_1$ (Eq.~(\ref{eq:sec4_onebin})). 
The quadratic estimator $\htau_{\ell m}$ depends on the fields $E_0$ and $E_1$ being correlated.
If the cross power spectrum $C_\ell^{E_0E_1}$ were zero, then the quadratic reconstruction would not be possible (the reconstruction noise
in Eq.~(\ref{eq:sec4_nl}) would be formally infinite).

\begin{figure}[htbp]
\begin{center}
\includegraphics[width=3.5in]{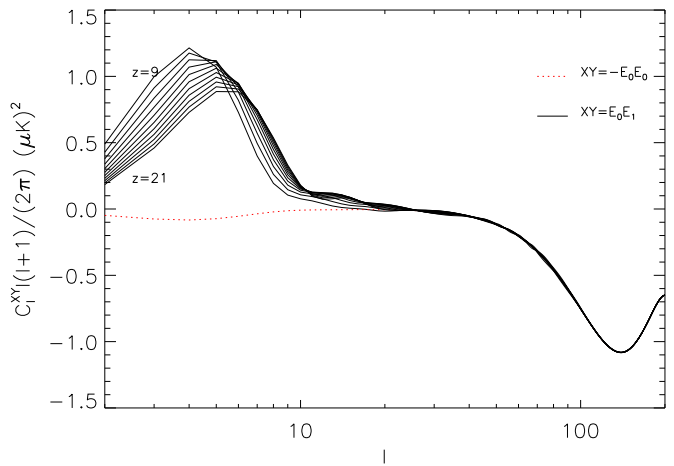}
\end{center}
\caption{Auto power spectrum $C_\ell^{E_0E_0}$ and cross power spectrum $C_\ell^{E_0E_1(z)}$ for varying redshift $9\le z\le 21$.
Here, $E_0$ is the E-mode without patchy reionization and $E_1(z)$ is the response field for $\tau$ fluctuations at redshift $z$, defined in Eq.~(\ref{eq:line_of_sight_qu1}).
On large scales, $C_\ell^{E_0E_1}$ is positive and dominated by the Thomson contribution to the response field.
On small scales, $C_\ell^{E_0E_1}$ is dominated by the screening contribution to the response field,
and $C_\ell^{E_0E_1} \approx -C_\ell^{E_0E_0}$.}
\label{fig:cle0e1}
\end{figure}

In Fig.~\ref{fig:cle0e1} we show the cross power spectrum $C_\ell^{E_0E_1}$, with the auto power spectrum $C_\ell^{E_0E_0}$
shown for comparison.
The cross power spectrum was calculated using a modified version of the CAMB code and the details of our calculation of $C_\ell^{E_0E_1}$ are given in App.~\ref{app:power_spectrum_calc}.

Note that $C_{\ell}^{E_0E_1}$ is positive at large scales due to Thomson scattering, while at small scales is negative due to the $e^{-\tau}$ screening.
To understand this effect one can think of the response field $E_1$ as the variation of the $E$ mode with $\tau$ fluctuations. Then, the addition of $\tau$ fluctuations generates more power in the reionization peak, but the $e^{-\tau}$ screening becomes larger too, so the amplitude of the acoustic peaks is suppressed.

\begin{figure*}[htbp]
\begin{center}
\begin{tabular}{c c c}
\includegraphics[width=2.2in]{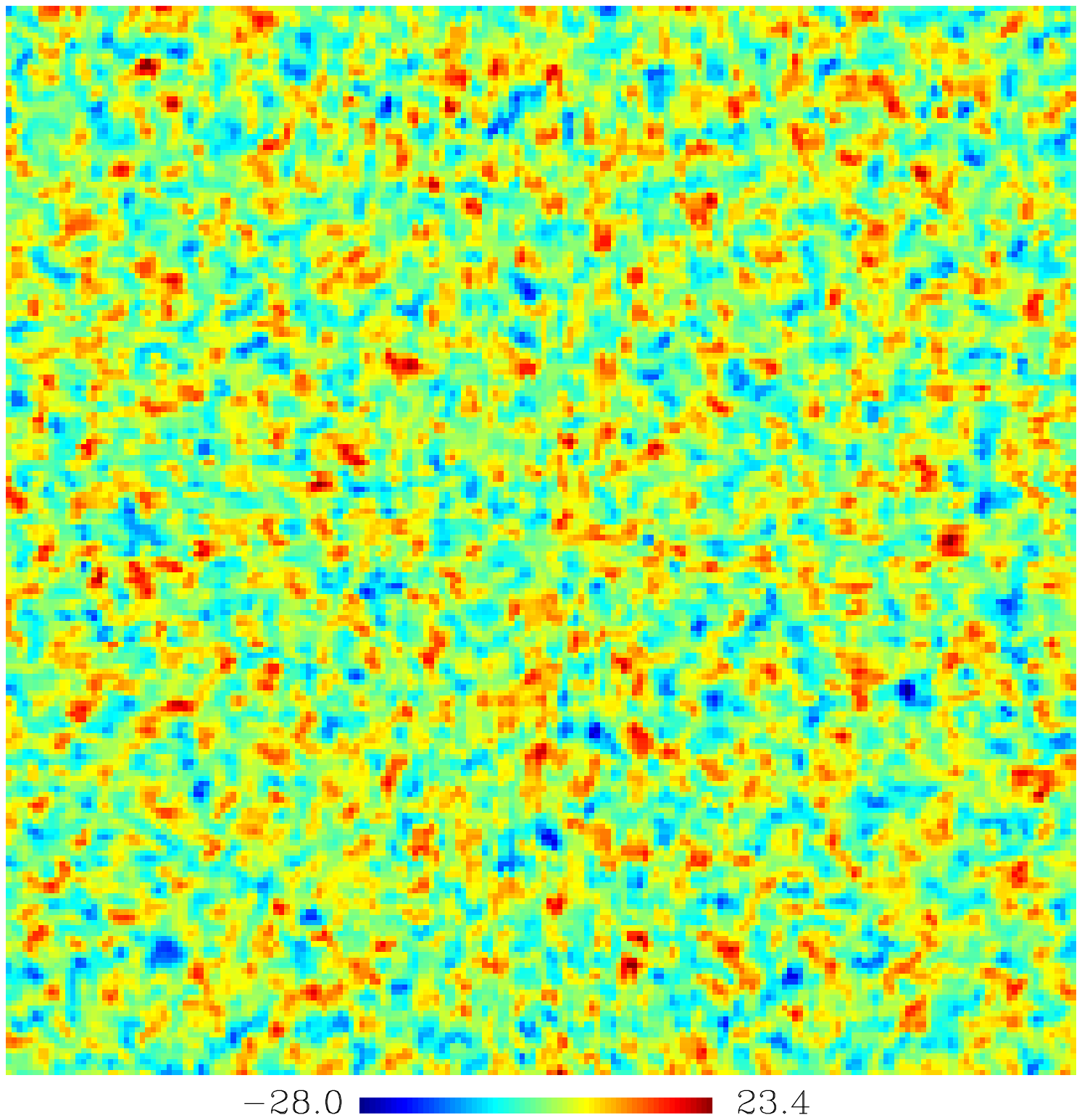} &
\includegraphics[width=2.2in]{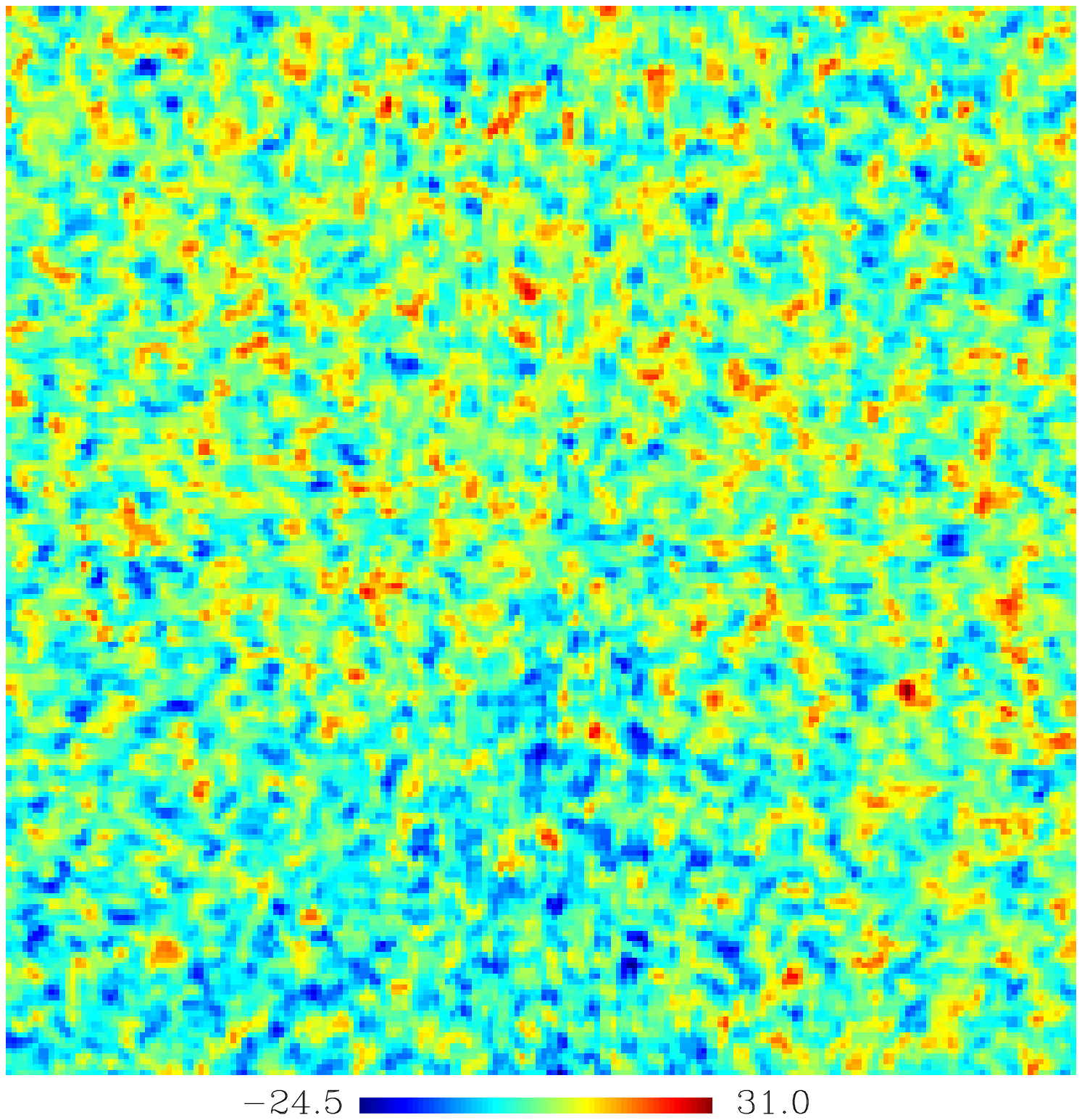} &
\includegraphics[width=2.2in]{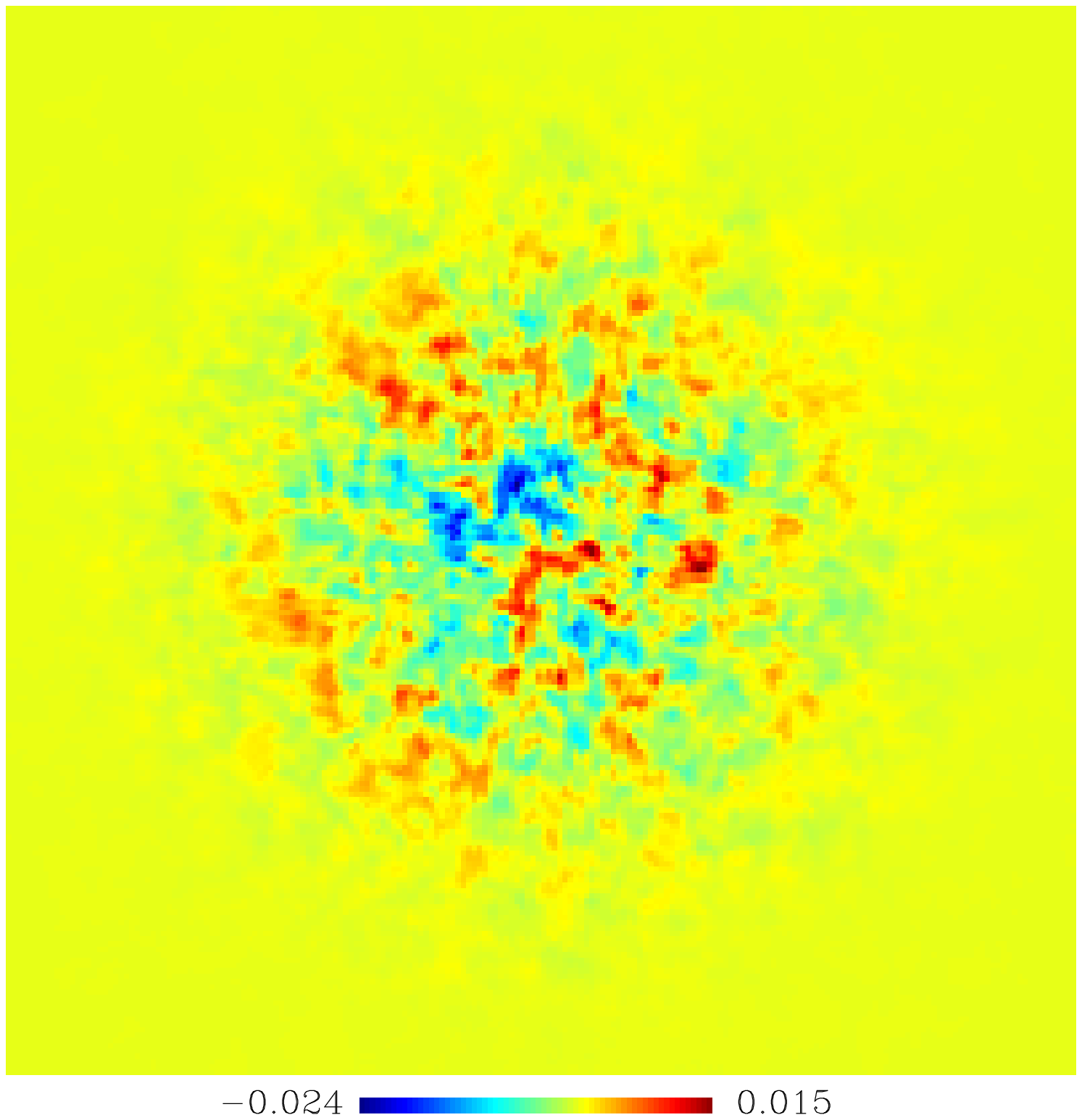} \\
\includegraphics[width=2.2in]{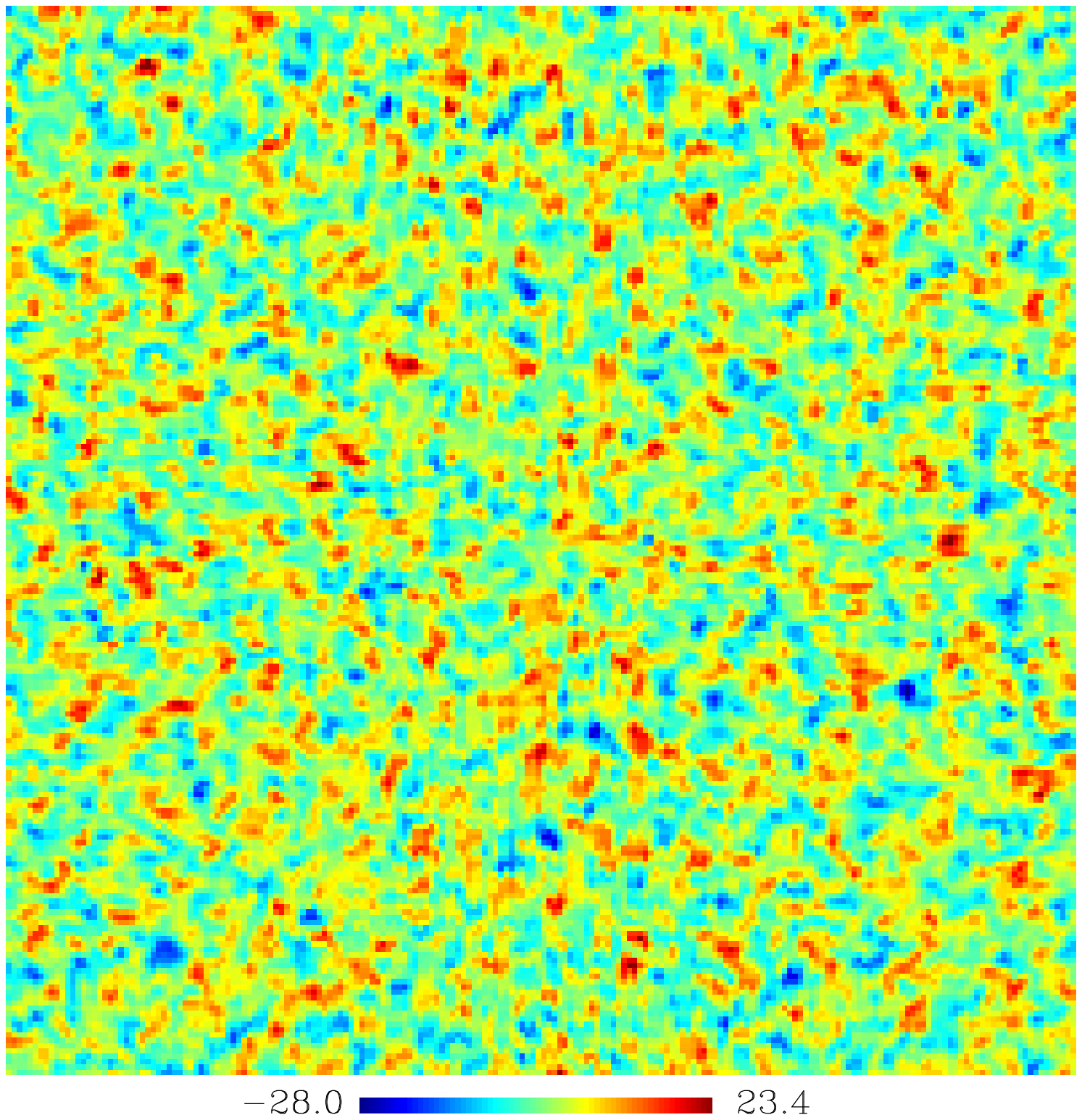} &
\includegraphics[width=2.2in]{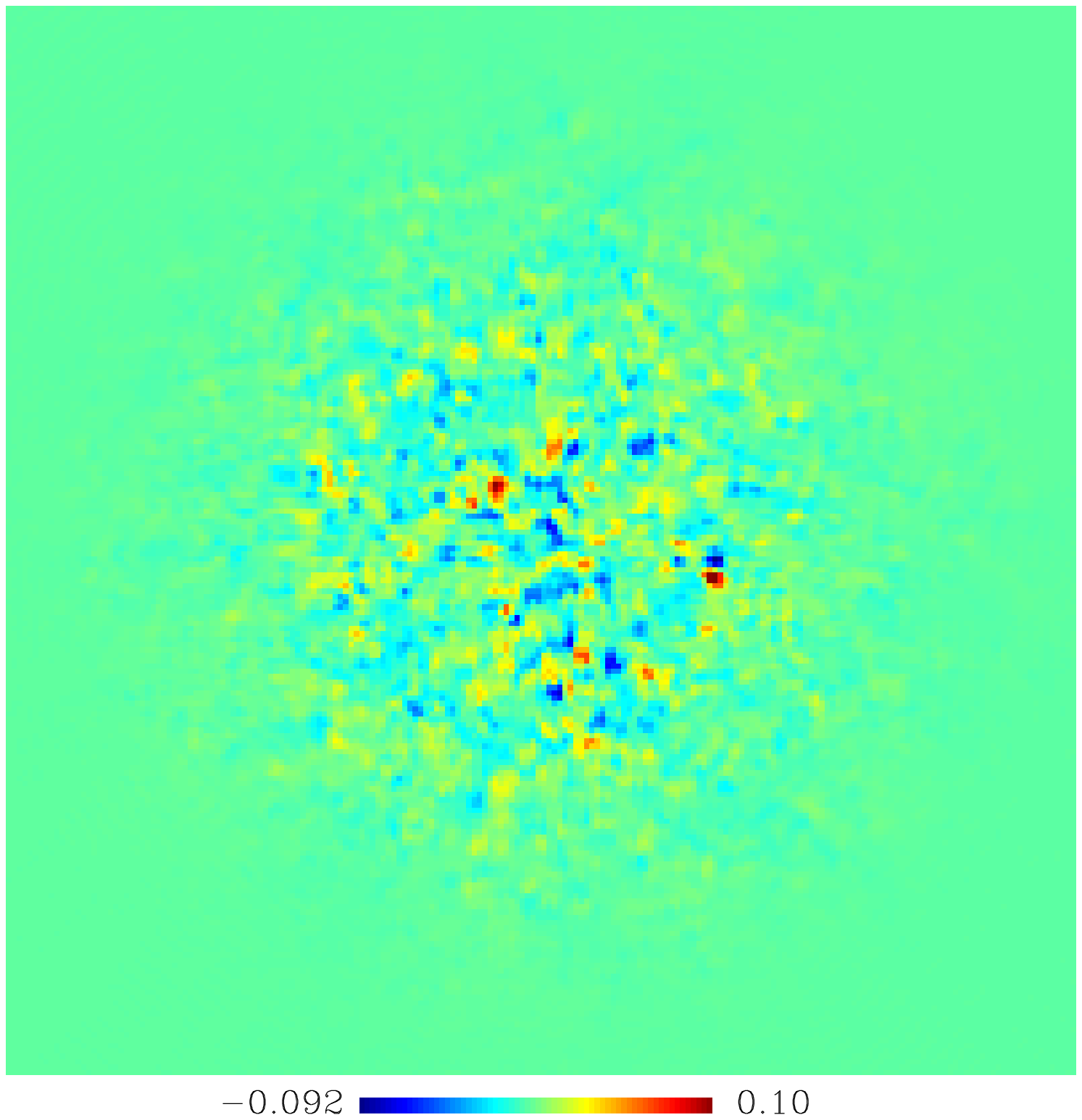} &
\includegraphics[width=2.2in]{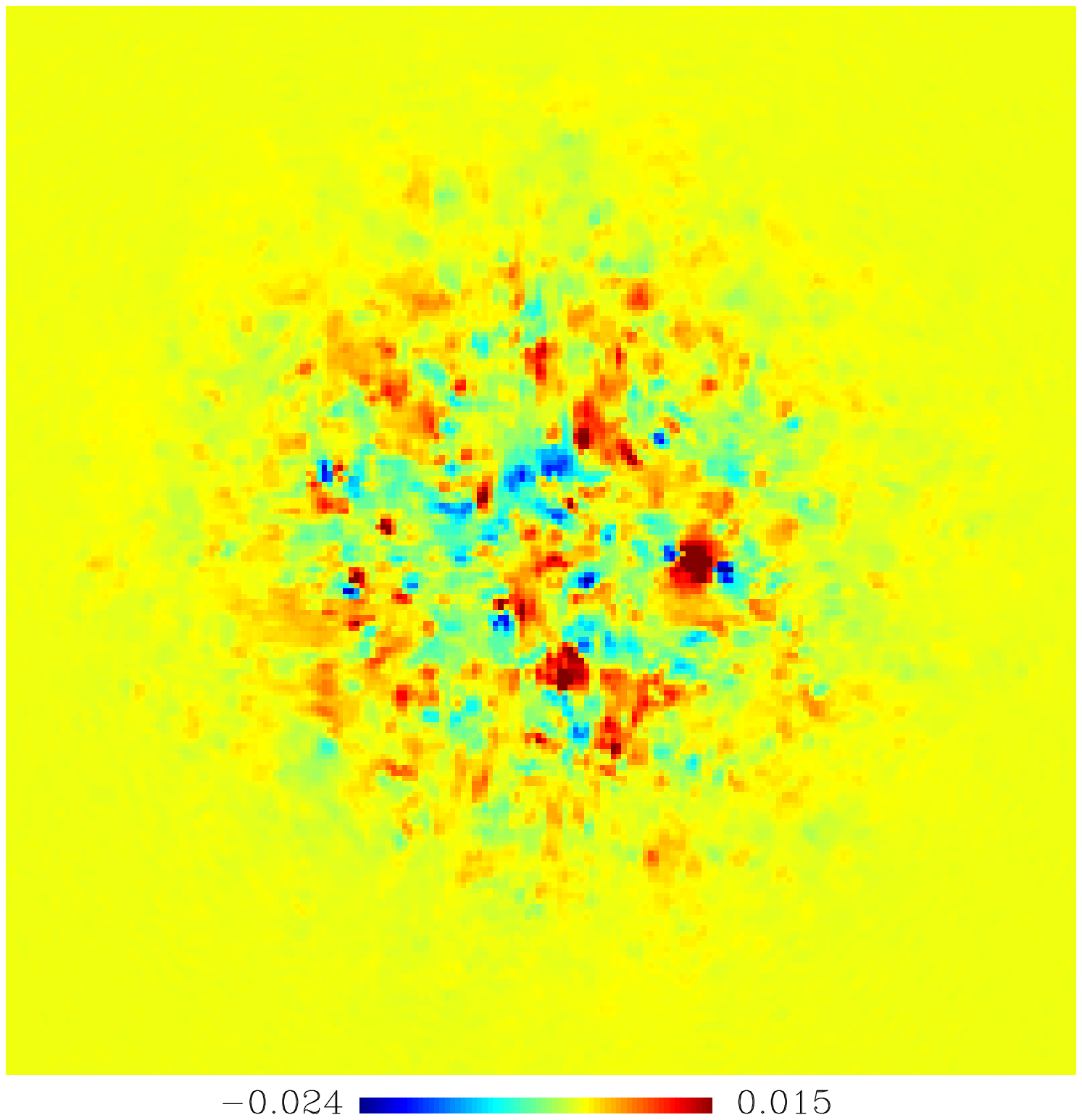}
\end{tabular}
\end{center}
\caption{An exaggerated example of the quadratic estimator $\htau_{\ell m}$ defined in \S\ref{sec:simple_estimator} assuming the constant quadrupole approximation.
{\em Top row} (left to right): the primary E-mode $E_0$, the response field $E_1$ and the $\Delta\tau$-field, on a 100 deg$^2$ patch of sky.
For visual purposes, the $\Delta\tau$-field has been multiplied by a Gaussian window function,
and we have artificially increased the signal-to-noise of the reconstruction by omitting the lensed B-mode and
assuming cosmic variance limited measurements to $\ellmax=2000$.
Note that $E_1 \approx -E_0$ on small scales, while the $E_1$-field has power added at large scales.
{\em Bottom row} (left to right): the $E$-mode and $B$-mode components of the total observed polarization (Eq.~(\ref{eq:sec4_onebin})), and the quadratic reconstruction
$\htau_{\ell m}$.
Note that $B$-modes appear where there are $\tau$ fluctuations.
In this figure, the units of the E-mode and B-mode polarization fields are in $\mu$K and $\Delta\tau$ is dimensionless.}
\label{fig:pedagogical}
\end{figure*}

In Fig.~\ref{fig:pedagogical} we show a toy example of our estimator $\htau_{\ell m}$ applied to a simulated polarization map (which was done using the HEALPIX package \cite{Gorski:2004by}),
with the signal-to-noise artificially increased by omitting the lensed B-mode from the simulation.  (Note that the inhomogeneous $\tau$
fluctuations will still generate a $B$-mode).
It is seen that the reconstruction $\htau_{\ell m}$ recovers the input field $\Delta\tau$ from the observed CMB, within some statistical noise associated with the reconstruction. 

The construction of the quadratic estimator $\htau_{\ell m}$ given in this section illustrates the qualitative
features of our method, but depends on an approximation that we would like to relax: the line-of-sight CMB quadrupole is 
constant throughout the patchy epoch.
Additionally, our estimator only uses the EB two-point function (Eq.~(\ref{eq:sec4_twopoint})).
In principle extra signal-to-noise
can be obtained by considering all cross-correlations of $\{T,E,B\}$.
For this reason, we will next consider a more general construction (\S\ref{sec:quad_est_realistic})
before presenting signal-to-noise forecasts in \S\ref{sec:forecasts}.

\section{Quadratic estimator for patchy reionization: complete treatment}
\label{sec:quad_est_realistic}

\subsection{Preliminaries}

In the previous section, we constructed a quadratic estimator $\htau_{\ell m}$ for the optical depth
anisotropy from patchy reionization, under the simplifying assumption of constant quadrupoles along the line-of-sight.
In this section, we will consider an extended version of this construction which does not rely on this approximation, and
which uses $TT$, $TE$, $EE$, and $TB$ cross-correlations in addition to $EB$.
This will introduce significant additional complexity, but in the end
we will show that a simple estimator which is very similar to the previous one
contains almost all of the signal-to-noise in practice.

As shown in \S\ref{sec:reionization_and_the_CMB}, 
if we divide the epoch of patchy reionization into redshift bins $\alpha$, then the observed CMB can be written:
\ba
T(\n) &=& T_0(\n)+\sum_{\alpha}\Delta\tau^{\alpha}(\n)T_1^{\alpha}(\n)  \label{eq:sec5_tqu} \\
(Q\pm iU)(\n) &=& (Q\pm iU)_0(\n)+\sum_{\alpha}\Delta\tau^{\alpha}(\n)(Q\pm iU)_1^{\alpha}(\n) \nn
\ea
All calculations in this section have been done using bins with $\Delta z=0.6$.

\begin{figure}[htbp]
\begin{center}
\includegraphics[width=3.5in]{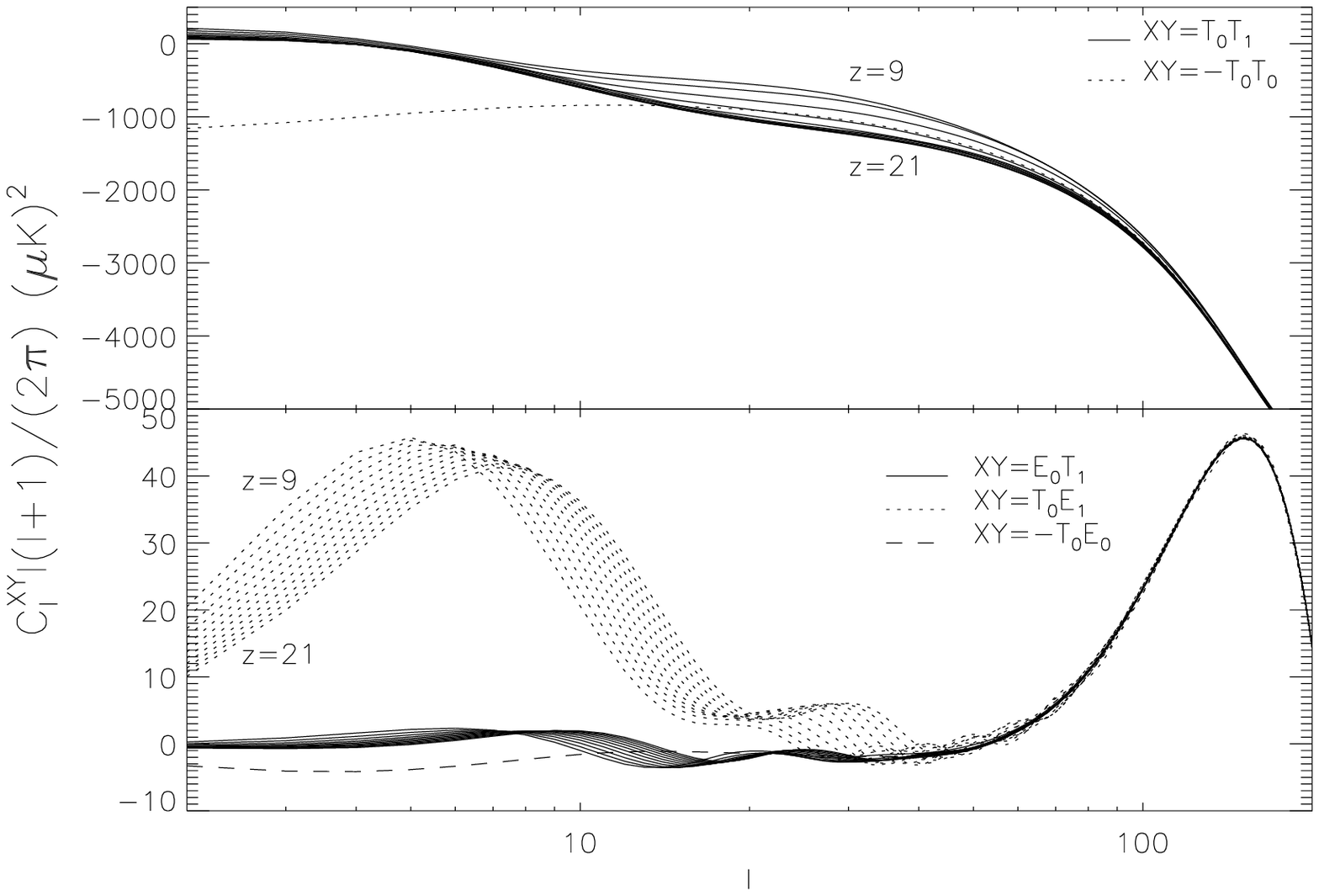}
\end{center}
\caption{Auto and cross power spectra $C_{\ell}^{T_0T_0}$, $C_{\ell}^{T_0E_0}$, $C_{\ell}^{T_0T_1}$, $C_{\ell}^{T_0E_1}$ and $C_{\ell}^{E_0T_1}$ which appear in the $\Gamma$
couplings (Eqs.~(\ref{eq:gamma_first})--(\ref{eq:gamma_last})) and in the quadratic estimator.
Notation follows \S\ref{sec:reionization_and_the_CMB}: $T_0,E_0$ are the temperature and E-mode polarization fields with patchy contribution omitted, and $T_1,E_1$ are the
response fields to $\tau$ fluctuations at redshift $z$.}
\label{fig:ClX0Y1}
\end{figure}

From Eq.~(\ref{eq:sec5_tqu}), it follows that the two-point function of the CMB, 
to first order in $\Delta\tau$, can be written: 
\ba
\left\langle a_{\ell_1m_1}^{X}a_{\ell_2m_2}^{Y}\right\rangle
&=&
(-1)^{m_1} C_{\ell_1}^{XY} \delta_{\ell_1\ell_2} \delta_{m_1,-m_2}  \label{eq:sec5_twopoint} \\
&& +
\sum_{\ell m\alpha}\Gamma^{XY (\tau_\alpha)}_{\ell_1\ell_2\ell}\wj{\ell_1}{\ell_2}{\ell}{m_1}{m_2}{m} (\Delta\tau^{\alpha})^{*}_{\ell m}, \nn
\ea
where the $\Gamma$ couplings are given by:    
\ba
\Gamma^{TT (\tau_\alpha)}_{\ell_1\ell_2\ell} &=& \left( C_{\ell_1}^{T_{0}T_{1} (\alpha)}+C_{\ell_2}^{T_{0}T_{1} (\alpha)} \right) J^{\ell_1 \ell_2 \ell}_{000} \label{eq:gamma_first}  \\
\Gamma^{EE (\tau_\alpha)}_{\ell_1\ell_2\ell} &=& {1\over 2} \left( C_{\ell_1}^{E_{0}E_{1} (\alpha)}+C_{\ell_2}^{E_{0}E_{1} (\alpha)} \right)\nn\\
&&\times \left[J^{\ell_1\ell_2\ell}_{-2,2,0}+J^{\ell_1\ell_2\ell}_{2,-2,0}\right]
\ea
\ba
\Gamma^{TE (\tau_\alpha)}_{\ell_1\ell_2\ell} &=& \frac{C_{\ell_1}^{T_{0}E_{1} (\alpha)}}{2}\left[J^{\ell_1\ell_2\ell}_{-2,2,0}+J^{\ell_1\ell_2\ell}_{2,-2,0}\right] \nn\\
&&+C_{\ell_2}^{T_{1}E_{0} (\alpha)}J^{\ell_1\ell_2\ell}_{000}\\
\Gamma^{TB (\tau_\alpha)}_{\ell_1\ell_2\ell} &=& {C_{\ell_1}^{T_{0}E_{1} (\alpha)}\over 2i}\left[J^{\ell_1\ell_2\ell}_{-2,2,0}-J^{\ell_1\ell_2\ell}_{2,-2,0}\right]\\
\Gamma^{EB (\tau_\alpha)}_{\ell_1\ell_2\ell} &=& {C_{\ell_1}^{E_{0}E_{1} (\alpha)}\over 2i}\left[J^{\ell_1\ell_2\ell}_{-2,2,0}-J^{\ell_1\ell_2\ell}_{2,-2,0}\right], \label{eq:gamma_last}
\ea
where we have defined
\be
J^{\ell_1\ell_2\ell}_{m_1m_2m}
=\sqrt{\frac{(2\ell_1+1)(2\ell_2+1)(2\ell+1)}{4\pi}}
\wj{\ell_1}{\ell_2}{\ell}{m_1}{m_2}{m}\,.
\ee
Note that the redshift dependence of the $\Gamma$ couplings enters through the cross power spectra $C_{\ell}^{X_0Y_1(\alpha)}$,
since the response field $Y_1$ is redshift-dependent.
In Fig.~\ref{fig:ClX0Y1} we show each of these cross power spectra as a function of $\ell$ and $z$.
The cross power spectra were calculated using a modified version of CAMB (App.~\ref{app:power_spectrum_calc}).

To get some intuition into these cross spectra, first note that for high $\ell$, the screening effect dominates and the response
fields satisfy $T_1\approx -T_0$, $E_1\approx -E_0$.  Each cross spectrum $C_\ell^{X_0Y_1}$ is approximately equal to $(-C_\ell^{X_0Y_0})$.
At low $\ell$, the Doppler contribution to $T_1$ and the Thomson contribution to $E_1$ become important, and one can see additional
features in the power spectra.
The largest ones are the low-$\ell$ bump in $C_\ell^{T_0E_1}$ (Fig.~\ref{fig:ClX0Y1}, lower panel),
which arises from correlation between the Doppler contribution to $T_0$ and the Thomson contribution to $E_1$,
and the bump in $C_\ell^{E_0E_1}$ (seen previously in Fig.~\ref{fig:cle0e1}) arising from correlation between the
reionization E-mode and the Thomson contribution to $E_1$.

\subsection{Principal component analysis}
\label{ssec:pca}

In this subsection, we will construct quadratic estimators which formally extract all the signal-to-noise, given
the TEB two-point function in Eq.~(\ref{eq:sec5_twopoint}).
This will generalize the single-field EB estimator from \S\ref{sec:simple_estimator}.
In the next subsection, we will show how to simplify the construction by constructing a simple estimator which
extracts nearly all the signal-to-noise in practice.

In App.~\ref{sec:optimal_estimator}, we show how to construct quadratic estimators from the TEB two-point function
(Eq.~(\ref{eq:sec5_twopoint})) using a principal component construction.
The key results can be summarized as follows.

There are $N$ ``eigenmode'' quadratic estimators $\hE^{(1)}_{\ell m}, \hE^{(2)}_{\ell m}, \ldots$ defined by:
\ba
\hE^{(i)}_{\ell m} =
{N_\ell^{\tau\tau (i)}\over 2} \sum_\alpha w^{(i)}_\ell(z^\alpha)
\sum_{\ell_1\ell_2}^{\ell_{\rm max}}\sum_{XYX'Y'm_1m_2}
\Gamma^{XY(\tau_\alpha)}_{\ell_1\ell_2\ell} \label{eq:eigenmode_htau} \nn\\
 \times
\wj{\ell_1}{\ell_2}{\ell}{m_1}{m_2}{m}
({\bf C}^{-1})^{XX'}_{\ell_1}a^{X'*}_{\ell_1m_1} ({\bf C}^{-1})^{YY'}_{\ell_2}a^{Y'*}_{\ell_2m_2}  
\ea
\ba
&&\frac{1}{N_\ell^{\tau\tau (i)}} =
\frac{1}{2(2\ell+1)} \sum_{\alpha\beta} w^{(i)}_\ell(z^\alpha) w^{(i)}_\ell(z^\beta) \label{eq:eigenmode_nl} \label{eq:sec5_eigen_nl} \\
&& \times \!\!\!\!\!
\sum_{XX'YY'\ell_1\ell_2}
({\bf C}^{-1})^{XX'}_{\ell_1}
({\bf C}^{-1})^{YY'}_{\ell_2}
\Gamma^{XY(\tau_\alpha)*}_{\ell_1\ell_2\ell}
\Gamma^{X'Y'(\tau_\beta)}_{\ell_1\ell_2\ell},  \nn
\ea
where $X,Y\in\{T,E,B\}$ and 
\be\label{eq:Cmatrix_def}
{\bf C}_\ell={\left(
                           \begin{array}{ccc}
        C_\ell^{TT}+N_\ell^{TT}& C_\ell^{TE}& 0\\
        C_\ell^{TE}& C_\ell^{EE}+N_\ell^{EE}& 0 \\
	0&0&C_\ell^{BB}+N_\ell^{BB}
                           \end{array}
                   \right)}
\ee

\begin{figure}[htbp]
\begin{center}
\includegraphics[width=3.6in]{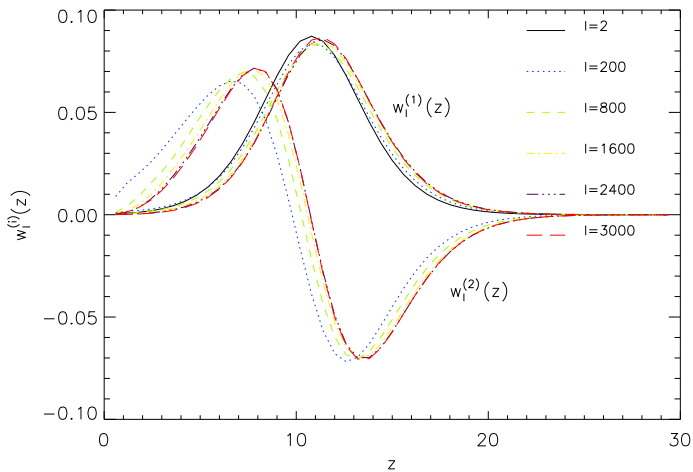}
\end{center}
\caption{Redshift weighting $w^{(i)}_\ell(z)$ for the first and second quadratic estimators $\hE^{(i)}_{\ell m}$ ($i=1,2$) defined in Eq.~(\ref{eq:eigenmode_htau}).
The redshift weightings have been normalized so that $\sum_{\alpha} w^{(1)}_\ell(z^\alpha) = 1$, 
and $\sum_{\alpha} (w^{(2)}_\ell(z^\alpha))^2 = 1/16$.  It is seen that the weightings are nearly independent of $\ell$.}
\label{fig:eigenvectors}
\end{figure}

The redshift weighting $w_\ell^{(i)}(z)$ which appears in the estimator is computed using a principal component analysis
described in App.~\ref{ssec:multiple_fields}.
In principle, the redshift weighting is $\ell$-dependent, but in practice the weighting is independent of $\ell$ to a good
approximation (Fig.~\ref{fig:eigenvectors}).
Note that the first eigenmode peaks at the redshift where $\overline x_e\approx 0.5$ in our fiducial model,
where we expect to get the largest contribution to the $\tau$-power spectrum.

The noise covariance of the quadratic estimators (generalizing Eq.~(\ref{eq:sec4_noise_covariance})
from the previous section) is given by:
\be
\langle \hE^{(i)*}_{\ell m} \hE^{(j)}_{\ell' m'} \rangle_{\rm noise}
=
N_\ell^{\tau\tau (i)} \delta_{ij} \delta_{\ell\ell'} \delta_{mm'},
\ee
i.e. the quantity $N_\ell^{\tau\tau (i)}$ defined in Eq.~(\ref{eq:sec5_eigen_nl})
is the reconstruction noise power spectrum of the $i$-th quadratic estimator and
the quadratic estimators are uncorrelated with each other.

\begin{figure}[htbp]
\begin{center}
\includegraphics[width=3.5in]{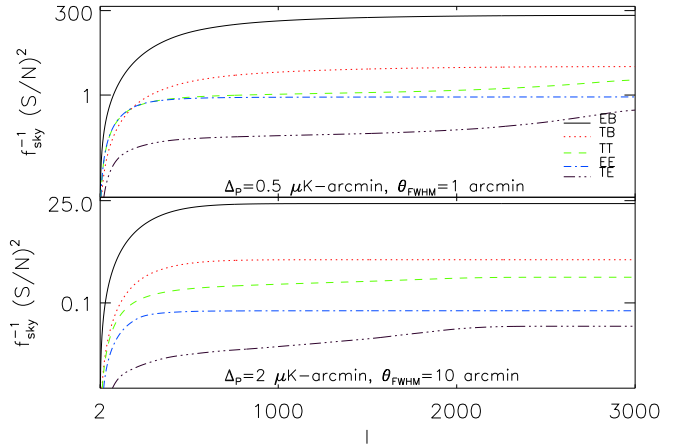}
\end{center}
\caption{Cumulative signal-to-noise versus maximum multipole $\ell$ in the $\tau$ reconstruction, for the eigenmode quadratic estimator (Eq.~(\ref{eq:eigenmode_htau})) split into contributions from
EB, TB, TT, EE, and TE cross-correlations.
In both panels, we take $\ellmax=3000$ and $\Delta_P=\sqrt{2}\Delta_T$.
In all cases, the EB pair has the dominant contribution to the signal-to-noise.}
\label{fig:ref_exp_A_D}
\end{figure}

The principal component analysis that is used to construct the estimators $\hE^{(i)}_{\ell m}$ also guarantees that
the estimators remain uncorrelated even in the presence of the $\Delta\tau$ signal.
We can therefore compute the total signal-to-noise of the detection of patchy reionization
by summing the signal-to-noise per $\ell$-mode for the $i$-th estimator (denoted by $\lambda_\ell^{(i)}$ and
defined precisely in Eq.~(\ref{eq:lambda_def})) over all values of $i$ and $\ell$:
\be
S/N = \left[\frac{\fsky}{2} \sum_{i\ell} (2\ell+1) (\lambda_{\ell}^{(i)})^2\right]^{1/2},  \label{eq:eigenmode_total_sn}
\ee
where $f_{\rm sky}$  refers to the observed fraction of the sky.

In Fig.~\ref{fig:ref_exp_A_D}, we show a signal-to-noise forecast for two experiments.
For the first experiment ($\Delta_P=2$ $\mu$K-arcmin, $\theta_{\rm FWHM}=10$ arcmin), we find $\fsky^{-1} (S/N)^2=21$.
This experiment is mainly limited by the beam size; at this resolution patchy reionization is detectable only by a large-$\fsky$
survey such as a future CMB polarization satellite.
(Some ideas for improving the $S/N$ are given in \S\ref{sec:discussion}.)
For the second experiment ($\Delta_P=0.5$ $\mu$K-arcmin, $\theta_{\rm FWHM}=1$ arcmin), we find $\fsky^{-1} (S/N)^2=217$.

\begin{figure}[htbp]
\begin{center}
\includegraphics[width=3.5in]{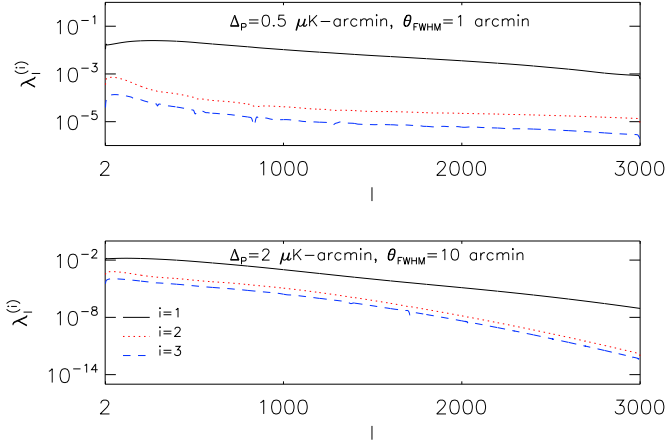}
\end{center}
\caption{Signal-to-noise per mode $\lambda_\ell^{(i)}$ for the first three ($i=1,2,3$) eigenmode quadratic estimators $\hE^{(i)}_{\ell m}$, keeping only the EB contribution.
It is seen that the first eigenmode contains nearly all the signal-to-noise.}
\label{fig:eigenvalues}
\end{figure}

In Fig.~\ref{fig:ref_exp_A_D} we also show the contribution to the total signal-to-noise from each of the 5 pairs of fields
TT, TE, EE, TB, EB in the quadratic estimator.  It is seen that almost all the signal-to-noise comes from the EB estimator. 
Also, note that the scale at which the signal-to-noise curves saturate ($\ell\approx 400$) is given by the scale at which the signal peaks.

Another way of splitting up the total signal-to-noise is to consider the contribution from 
each of the quadratic estimators.
When this is done, it is seen (Fig.~\ref{fig:eigenvalues}) that the first estimator $\hE^{(1)}_{\ell m}$ contains almost all
the signal-to-noise.
Intuitively, this means that our reconstruction of patchy reionization has ``the degrees of freedom of a single 2D field''.
(The opposite limiting situation, i.e. many quadratic estimators with roughly the same signal-to-noise but orthogonal
redshift weightings, would correspond to reconstructing the degrees of freedom of a 3D field.)

\begin{figure}[htbp]
\begin{center}
\includegraphics[width=3.3in]{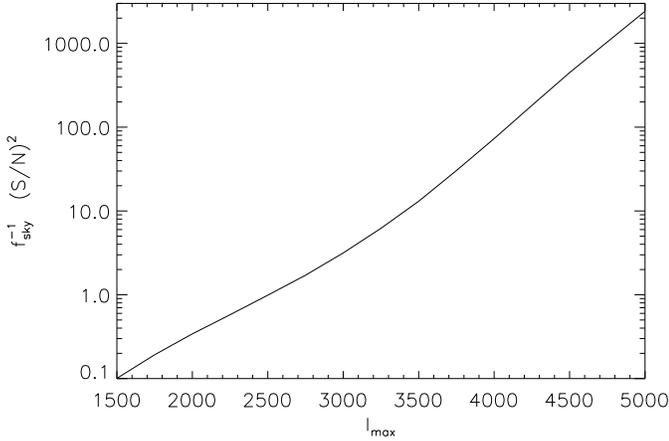}
\end{center}
\caption{Total signal-to-noise for the $TT$ quadratic estimator, assuming cosmic variance limited temperature measurements up to some maximum multipole $\ellmax$,
and treating the lensing contribution to the temperature power spectrum as a source of noise.  For $\ellmax\gtrsim 4000$, the signal-to-noise rises sharply but in
practice the measurement is likely to be contaminated by non-Gaussian secondary anisotropies.}
\label{fig:StoN_TT_noiseless}
\end{figure}

An interesting result emerges when we extend the analysis to smaller scales (i.e. $\ell_{\rm max} > 3000$). At those scales, the TT estimator acquires large signal-to-noise. 
Fig.~\ref{fig:StoN_TT_noiseless} shows the signal-to-noise coming from the TT pair as a function of $\ell_{\rm max}$. 
In principle, at smaller scales ($\ell_{\rm max}\sim 5000$), the TT estimator is more sensitive than EB.
However, it is unclear whether other secondaries such as point sources, kinetic SZ and thermal SZ effects will permit 
a clean measurement of patchy reionization on these scales.  We defer a complete treatment of this issue to future work.

\subsection{A simple estimator which contains all the S/N}
\label{ssec:simple_estimator}

In the previous subsection, we constructed a complete set of eigenmode estimators which formally extract all the signal-to-noise
given the TEB two-point function in Eq.~(\ref{eq:sec5_twopoint}).
However, the conclusion of this analysis was that almost all the signal-to-noise was contained
in the first principal component $\hE_{\ell m}^{(1)}$.
Intuitively, this means that all the redshift bins are so highly correlated that the choice of redshift weighting
is not important.
Furthermore, we found that all the signal-to-noise was contained in the EB estimator.

These observations motivate the following construction.
We fix a redshift bin $\mu$
and construct an estimator $\htau_{\ell m}^{(\mu)}$ by simply repeating the ``toy''
example from \S\ref{sec:simple_estimator}, using the $\Gamma$ couplings corresponding to bin $\mu$:
\ba
\hat{\tau}_{\ell m}^{(\mu)} &=& N_\ell^{\tau\tau (\mu)} \sum_{\ell_1m_1\ell_2m_2}\Gamma_{\ell_1\ell_2\ell}^{EB (\tau_\mu)}\wj{\ell_1}{\ell_2}{\ell}{m_1}{m_2}{m} \nn\\
&& \times {a_{\ell_1m_1}^{E*}a_{\ell_2m_2}^{B*}\over (C_{\ell_1}^{EE}+N_{\ell_1}^{EE})(C_{\ell_2}^{BB}+N_{\ell_2}^{BB})}\label{eq:taueff_def}
\ea
where:
\be\label{eq:eff_noise}
N_{\ell}^{\tau\tau (\mu)}= \left[ {1\over 2\ell+1}\sum_{\ell_1\ell_2}{ |\Gamma_{\ell_1\ell_2\ell}^{EB (\tau_\mu)}|^2 \over(C_{\ell_1}^{EE}+N_{\ell_1}^{EE})(C_{\ell_2}^{BB}+N_{\ell_2}^{BB})} \right]^{-1} 
\ee
We will now compute the signal-to-noise of this estimator in a way which does not rely on the constant quadrupole approximation from \S\ref{sec:simple_estimator},
and show that the signal-to-noise is nearly the same as the optimal estimator from the previous subsection.

Without the constant quadrupole approximation, the expectation value of the reconstruction has a nontrivial redshift dependence:
\be
\left\langle \htau_{\ell m}^{(\mu)} \right\rangle = 
\sum_\alpha (R_\ell^{(\mu)}(z^\alpha)) (\Delta\tau^{\alpha})_{\ell m},
\ee
where the redshift response $R_\ell^{(\mu)}(z^\alpha)$ is given by:
\be
R_\ell^{(\mu)}(z^\alpha) = \frac{N_\ell^{\tau\tau (\mu)}}{2\ell+1} 
\sum_{\ell_1\ell_2}{\Gamma_{\ell_1\ell_2\ell}^{EB (\tau_\mu)}\Gamma_{\ell_1\ell_2\ell}^{EB(\tau_\alpha)*}\over(C_{\ell_1}^{EE}+N_{\ell_1}^{EE})(C_{\ell_2}^{BB}+N_{\ell_2}^{BB})}   \label{eq:simple_rr}
\ee
The total signal-to-noise is given by
\be\label{eq:simple_sn}
S/N = \left[\frac{\fsky}{2} \sum_\ell (2\ell+1) \left( \frac{C_\ell^{\tau\tau\rm (eff)}}{N_\ell^{\tau\tau (\mu)}} \right)^2\right]^{1/2},  
\ee
where
\be
C_\ell^{\tau\tau\rm (eff)} = \sum_\alpha \left(R_\ell^{(\mu)}(z^\alpha)\right)^2\, C_\ell^{\tau_\alpha\tau_\alpha}
\ee

\begin{figure}[h]
\begin{center}
\includegraphics[width=3.6in]{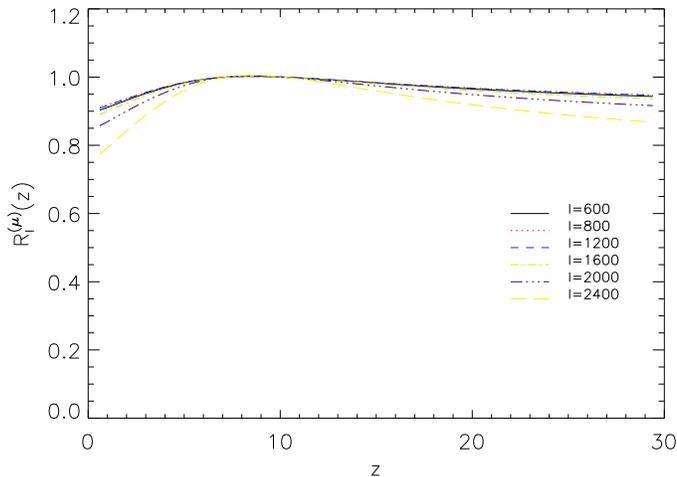}
\end{center}
\caption{Redshift response $R_\ell^{(\mu)}(z)$ for the simple estimator, defined in Eq.~(\ref{eq:simple_rr}), showing a near-unit response ($R_\ell^{(\mu)}(z)\approx 1$) for all redshifts.
This plot was made for a redshift bin $\mu$ that corresponds to $z=10.2$.}
\label{fig:mean_response_specific_bin}
\end{figure}

If we consider a survey with 
$\Delta_P=0.1$ $\mu$K-arcmin, $\theta_{\rm FWHM}=1$ arcmin, $\ellmax=3000$,
and take the redshift bin $\mu$ which corresponds to $z=10.2$,
we find the following numerical results.
(The qualitative conclusions do not depend on these choices.)
The signal-to-noise for the simple estimator (Eq.~(\ref{eq:simple_sn})) is nearly the same ($\sim$99.9\%)
as the principal component estimator (Eq.~(\ref{eq:eigenmode_total_sn})).
The redshift response $R_\ell(z)$ is $\approx$1, i.e. the estimator $\htau^{(\mu)}_{\ell m}$
simply reconstructs the sum of the $\tau$ fluctuations over all redshift bins (Fig.~(\ref{fig:mean_response_specific_bin})).
Intuitively, this is because the redshift bins are so highly correlated that the estimator which is optimized
for $\tau$ fluctuations in bin $\mu$ will pick up the fluctuations from every bin.
The near-unit redshift response also implies that
\be
C_\ell^{\tau\tau ({\rm eff})} \approx \sum_\alpha C_\ell^{\tau_\alpha\tau_\alpha}  \label{eq:cltaueff_approx}
\ee
Numerically, we find that this agreement is better than $1\%$.

Let us conclude this section by discussing the issue of model uncertainty.
Our construction of $\htau_{\ell m}$ depends on knowing the cross spectrum $C_\ell^{E_0E_1(z)}$,
which depends on cosmological parameters.
If we have imprecise knowledge of the cosmology, does this affect the estimator in a significant way
(for example, by changing the normalization)?

We expect that the most important parameter dependence will be the $\tau$ dependence of the cross
spectrum $C_\ell^{E_0E_1}$.  (The dependence of the reionization E-mode on geometric distances, for
example, should be much weaker than the $\tau$ dependence.)
At low $\ell$, the cross spectrum is roughly proportional to $\tau$,
because the largest contribution is from cross-correlating the patchy Thomson signal with the
primary reionization E-mode.  
At high $\ell$, the cross spectrum is roughly independent of $\tau$, provided that $\tau$
is varied with the heights of the acoustic peaks held fixed, because $C_\ell^{E_0E_1}\approx -C_\ell^{E_0E_0}$.

The forecasts in \S\ref{sec:forecasts} will show that we expect a $\lesssim 10\sigma$ detection for futuristic
sensitivity levels.  The uncertainty $\sigma(\tau)$ would have to be $\gtrsim 0.01$ (i.e. a fractional
uncertainty of 10\%) in order to bias this measurement.  The forecasted uncertainty from Planck is
$\sigma(\tau) \approx 0.0045$ \cite{Smith:2006nk}, so parameter uncertainty is unlikely to be an issue.

In conclusion, the principal component construction from the previous subsection
should be regarded as a proof that the simple estimator presented here extracts all
the information.
For practical analysis, the principal component analysis is not necessary and one can use the simple
estimator constructed in Eq.~(\ref{eq:taueff_def}).
For forecasting, one can compute a noise power spectrum for the simple estimator using Eq.~(\ref{eq:eff_noise}),
and the signal power spectrum is given to an excellent approximation by Eq.~(\ref{eq:cltaueff_approx}).
Uncertainty in cosmological parameters is unlikely to be a complicating factor.

\section{Simulations}
\label{sec:simulation}

In this section, we will demonstrate our quadratic estimator in Monte Carlo simulations which include 
the patchy reionization contribution to the CMB.
Fully realistic simulations are outside the scope of this paper; we will make some simplifying assumptions,
deferring a more complete study to future work currently in preparation \footnote{C. Dvorkin, W. Hu, and K. M. Smith, in preparation.}.
Our most critical simplifying assumption will be to simulate the lensing component of the B-mode
as if it were a Gaussian field with the power spectrum of the true lensed B-mode.

In fact, the true statistics of the lensed B-mode are non-Gaussian and this is a significant
concern for our quadratic estimator.  Our power spectrum estimator $\hC_\ell^{\tau\tau}$ is
a four-point estimator in the CMB, and gravitational lensing generates a large four-point 
function \cite{Zaldarriaga:2000ud,Hu:2001fa,Okamoto:2002ik},
which could potentially be a source of bias.
One promising solution to this problem is to apply delensing algorithms \cite{Seljak:2003pn}: 
if the four-point function induced by gravitational lensing is large enough to be detected statistically, then
the level of lensing contamination can be reduced by partially reconstructing the lensing potential.
This procedure can be iterated \cite{Hirata:2003ka} to obtain an optimal reconstruction which minimizes the level of the
residual four-point function.
Preliminary results from work in preparation indicate that an analysis procedure which combines delensing with the reionization
estimator from this paper, by solving simultaneously for the $\phi$ and $\Delta\tau$ fields, is a promising approach, but this
is outside the scope of this paper and so we use Gaussian simulations here.

\subsection{Practical estimator}

The harmonic-space form of the estimator given in \S\ref{sec:simple_estimator}:
\ba
\htau_{\ell m} &=& N_\ell^{\tau\tau} \sum_{\ell_1m_1\ell_2m_2} \frac{\Gamma^{EB (\tau)}_{\ell_1\ell_2\ell}}{(C_{\ell_1}^{EE}+N_{\ell_1}^{EE})(C_{\ell_2}^{BB}+N_{\ell_2}^{BB})} \nn\\
&& \times \wj{\ell_1}{\ell_2}{\ell}{m_1}{m_2}{m} a^{E*}_{\ell_1m_1} a^{B*}_{\ell_2m_2} \label{eq:tauhat_sec6} 
\ea
\be
N_\ell^{\tau\tau}= \left[ \frac{1}{2\ell+1}\sum_{\ell_1\ell_2} 
\frac{|\Gamma^{EB (\tau)}_{\ell_1\ell_2\ell}|^2}{(C_{\ell_1}^{EE}+N_{\ell_1}^{EE})(C_{\ell_2}^{BB}+N_{\ell_2}^{BB})} \right]^{-1}  \label{eq:nl_sec6}
\ee
is convenient for forecasting signal-to-noise, but too slow for practical evaluation: the computational cost is
$\bigoh(\ellmax^5)$, which is prohibitively expensive for $\ellmax \gtrsim 200$.

Fortunately, Eq.~(\ref{eq:tauhat_sec6}) can be written in a fast position-space form.
First we define spin-$2$ fields $\chi$ and $\psi$:
\ba
\chi(\n) &=& \sum_{\ell m} \left( \frac{i}{C_\ell^{BB}+N_\ell^{BB}} \right) a_{\ell m}^B ({}_2Y_{\ell m}(\n))  \\
\psi(\n) &=& \sum_{\ell m} \left( \frac{C_\ell^{E_0E_1}}{C_\ell^{EE}+N_\ell^{EE}} \right) a_{\ell m}^E ({}_2Y_{\ell m}(\n))
\ea
Then the estimator $\htau_{\ell m}$ is given in position-space form by
\be
\htau_{\ell m} = \frac{N_\ell^{\tau\tau}}{2} \int d^2\n\, Y_{\ell m}^*(\n)[ \chi(\n)^* \psi(\n) + \chi(\n) \psi(\n)^* ]
\ee
In this form, the computational cost of the estimator is $\bigoh(\ellmax^3)$, using a fast isolatitude spherical transform.

It is also convenient to have a fast position-space form for computing the noise power spectrum $N_\ell^{\tau\tau}$.
If we define correlation functions $\zeta^\pm_\chi$ and $\zeta^\pm_\psi$ by
\ba
\zeta^\pm_\chi &=& \pm \sum_\ell \left( \frac{2\ell+1}{4\pi} \right) 
                      \frac{1}{C_\ell^{BB} + N_\ell^{BB}} d^\ell_{2,\pm 2}(\theta) \label{eq:zetachi_def} \\
\zeta^\pm_\psi &=& \sum_\ell \left( \frac{2\ell+1}{4\pi} \right) 
                        \frac{(C_\ell^{E_0E_1})^2}{C_\ell^{EE} + N_\ell^{EE}} d^\ell_{2,\pm 2}(\theta),
\ea
then $N_\ell^{\tau\tau}$ is given by
\be
(N_\ell^{\tau\tau})^{-1} = \pi \int_{-1}^1 d(\cos\theta) d^\ell_{00}(\theta) 
  [\zeta^+_\chi(\theta) \zeta^+_\psi(\theta) + \zeta^-_\chi(\theta) \zeta^-_\psi(\theta)]  \label{eq:fast_nl},
\ee
where $d^\ell_{2,\pm 2}$ and $d^\ell_{00}$ are reduced Wigner $D$-functions.
This form of $N_\ell^{\tau\tau}$ is faster ($\bigoh(\ellmax^2)$ vs. $\bigoh(\ellmax^3)$) than the harmonic-space form
given previously in Eq.~(\ref{eq:nl_sec6}).

Before presenting Monte Carlo results, there is one more ingredient.
The most straightforward way to estimate a $\tau$-power spectrum would be to use the power spectrum estimator:
\be
\hC_\ell^{\tau\tau} = \left( \frac{1}{2\ell+1} \sum_m \htau_{\ell m}^* \htau_{\ell m} \right) - N_\ell^{\tau\tau},   \label{eq:hc_bad}
\ee
where $\htau_{\ell m}$ is the quadratic estimator.
Note that this definition of the estimator $\hC_\ell^{\tau\tau}$ includes subtraction of the noise bias $N_\ell^{\tau\tau}$.
In turn, $N_\ell^{\tau\tau}$ is a function of the CMB power spectra $C_\ell^{EE}$ and $C_\ell^{BB}$.
One can ask, if the observed $C_\ell$'s in our sky can differ (within cosmic variance) from the model $C_\ell$'s, 
should we compute the noise bias using the model $C_\ell$'s (as we have assumed when defining $\hC_\ell^{\tau\tau}$ 
in Eq.~(\ref{eq:hc_bad})) or the observed $C_\ell$'s?

In fact, simulations show that the variance of the estimator $\hC_\ell^{\tau\tau}$ is much smaller if the noise
bias $N_\ell^{\tau\tau}$ is recomputed in each Monte Carlo realization, using the observed $C_\ell$'s in the realization.
Intuitively, this reduction in variance can be thought of as subtracting scatter in the noise power spectrum $N_\ell^{\tau\tau}$ 
which arises because the CMB $C_\ell$'s will fluctuate (within cosmic variance) from realization to realization.
Formally, we define an improved power spectrum estimator $\hC_\ell^\imp$ in the following way:
\be
\hC_\ell^\imp = \left( \frac{1}{2\ell+1} \sum_m \htau_{\ell m}^* \htau_{\ell m} \right) - \hN_\ell^{\tau\tau},  \label{eq:hc_improved}
\ee
where the estimator $\hN_\ell^{\tau\tau}$ is defined by:
\ba
\hN_\ell^{\tau\tau} &=& 
\frac{(N_\ell^{\tau\tau})^2}{2\ell+1}\sum_{\ell_1\ell_2} 
   \frac{|\Gamma^{EB (\tau)}_{\ell_1\ell_2\ell}|^2}{(C_{\ell_1}^{EE}+N_{\ell_1}^{EE})^2(C_{\ell_2}^{BB}+N_{\ell_2}^{BB})^2} \nn \\
&& \times \sum_{m_1m_2} \frac{a_{\ell_1m_1}^{E*} a_{\ell_1m_1}^E a_{\ell_2m_2}^{B*} a_{\ell_2m_2}^B}{(2\ell_1+1)(2\ell_2+1)} \label{eq:nlhat_def}
\ea
Note that a fast position-space algorithm for computing $\hN_\ell^{\tau\tau}$, analogous to Eqs.~(\ref{eq:zetachi_def})--(\ref{eq:fast_nl}),
is given as follows:
\ba
\zeta^\pm_\chi &=& \pm \frac{1}{4\pi} \sum_{\ell m} \frac{a_{\ell m}^{B*} a_{\ell m}^B}{(C_\ell^{BB} + N_\ell^{BB})^2} 
                      d^\ell_{2,\pm 2}(\theta) \\
\zeta^\pm_\psi &=& \frac{1}{4\pi} \sum_{\ell m}
                 \frac{(C_\ell^{E_0E_1})^2 a^{E*}_{\ell m} a^E_{\ell m}}{(C_\ell^{EE} + N_\ell^{EE})^2} d^\ell_{2,\pm 2}(\theta) \nn 
\ea
\be
\hN_\ell^{\tau\tau} = \pi (N_\ell^{\tau\tau})^2 \int_{-1}^1 d(\cos\theta) d^\ell_{00}(\theta) 
  [\zeta^+_\chi(\theta) \zeta^+_\psi(\theta) + \zeta^-_\chi(\theta) \zeta^-_\psi(\theta)]  \nn
\ee

In addition to the improvement in variance, another advantage of $\hC_\ell^\imp$ 
is that the noise bias subtraction does not depend on having precise knowledge of the fiducial model.
The model $C_\ell$'s are only used in the estimator (via the $(C_\ell+N_\ell)$ denominators in 
Eqs.~(\ref{eq:tauhat_sec6}),~(\ref{eq:nl_sec6}) and ~(\ref{eq:nlhat_def})) for weighting purposes.
The improved estimator $\hC_\ell^\imp$ only depends weakly on the choice of model $C_\ell$'s and, more importantly, is not biased if the wrong model is used.

We have defined $\hC_\ell^\imp$ in such a way that the expectation value $\langle \hC_\ell^\imp\rangle$ is always 
zero when taken over any ensemble of Gaussian CMB realizations (even if the CMB power spectra differ from the fiducial ones), 
but becomes nonzero in the presence of the non-Gaussian patchy reionization signal.
This is particularly important for the reionization E-mode bump at low $\ell$, where the model $C_\ell$'s 
may have large uncertainties due to the number of reionization parameters that must be fit from the reionization bump itself.

\subsection{Monte Carlo pipeline}

The first step in our Monte Carlo pipeline is to simulate the observed CMB polarization $[Q\pm iU]_{\rm obs}$.
Our simulation procedure includes several simplifying assumptions as we now describe.
We simulate $\Delta\tau$ as a Gaussian field with power spectrum given by Eq.~(\ref{eq:cltau}), rather than simulating bubble formation during reionization.
We simulate the primary E-mode $E_0$ and response field $E_1$ as correlated Gaussian fields, making the constant quadrupole assumption
(from \S\ref{sec:simple_estimator}) so that $E_1$ can be treated as a 2D field.
In a more complete treatment we would include the redshift dependence of $E_1$ as described in \S\ref{ssec:pca}.
As discussed above, we simulate the lensed B-mode $B_{\rm lensed}$ as if it were a Gaussian field with the correct power spectrum,
rather than including the true non-Gaussian statistics.
We then compute the observed polarization:
\ba
[Q\pm iU]_{\rm obs}(\n) &=& [Q\pm iU]_0(\n) + [Q\pm iU]_{\rm lensed}(\n) \nn \\
                        && \hskip 0.2in + (\Delta\tau)(\n) [Q\pm iU]_1(\n)
\ea
Note that these simulations do not include noise. We will present Monte Carlo results for an ensemble of cosmic variance
limited simulations to $\ellmax=2000$.

We next apply the power spectrum estimator $\hC_\ell^\imp$ defined in Eq.~(\ref{eq:hc_improved}) 
to the observed polarization in each realization.

\begin{figure}[htbp]
\centerline{\epsfxsize=3.2truein\epsffile[30 200 305 420]{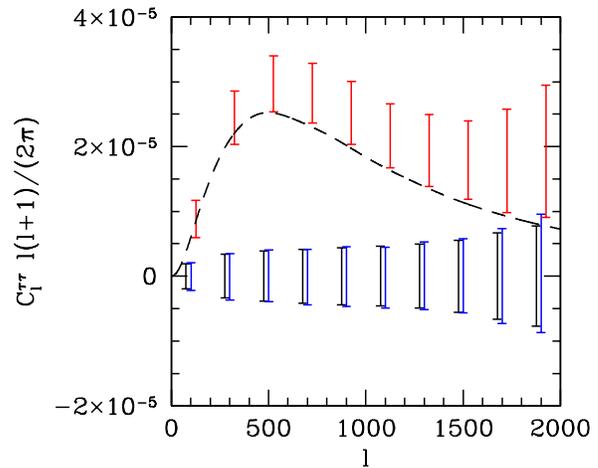}}
\caption{Comparison between forecasted and simulated $\tau$-power spectrum errors,
in bandpowers with $\Delta\ell=200$ and assuming cosmic variance limited measurements up to $\ellmax=2000$.
{\em Black/left:} Forecasted power spectrum errors, using the signal-to-noise calculations from \S\ref{sec:quad_est_realistic}.
{\em Blue/middle:} Monte Carlo power spectrum errors using the estimator $\hC_\ell^\imp$ (Eq.~(\ref{eq:hc_improved})),
in an ensemble of simulations without patchy reionization ($\tau=0.084, \Delta_y=0$).
{\em Red/right:} Monte Carlo power spectrum errors in the fiducial patchy reionization model
($\tau=0.084, \Delta_y=19.0$).
The dashed curve is the fiducial power spectrum $C_\ell^{\tau\tau}$.}
\label{fig:mc_many}
\end{figure}

Results from $1000$ Monte Carlo simulations are shown in Fig.~\ref{fig:mc_many}.
We bin the estimated power spectrum in bands with $\Delta\ell=200$ and we assign a bandpower that is the average of $\ell(\ell+1)\hC_\ell^\imp/(2\pi)$ over each band.
In an ensemble of simulations with {\em homogeneous} reionization, it is seen that the estimated $\tau$-power spectrum
has mean zero, and bandpower errors which are consistent with calculations in \S\ref{sec:quad_est_realistic}.
When patchy reionization is included in the simulations, the estimated $\tau$-power spectrum acquires a nonzero
expectation value which matches the fiducial power spectrum $C_\ell^{\tau\tau}$ at low $\ell$, but is biased
high at high $\ell$.

This power spectrum bias is a known phenomenon in the context of lens reconstruction, where the expectation
value of the estimator $\hC_\ell^{\phi\phi}$ is linear in the power spectrum $C_\ell^{\phi\phi}$ of the lens
potential, but the matrix which relates the two is not the identity matrix \cite{Cooray:2002py}: an additional
term appears when one writes out all possible contractions using Wick's theorem.
The bias can be absorbed into the normalization of the power spectrum estimator,
either by computing the relevant matrix analytically, or using an iterative approach as
suggested in \cite{Cooray:2002py}.
However, we will not characterize the bias in detail in this paper; our objective here is simply to demonstrate
that our signal-to-noise forecasts in \S\ref{sec:quad_est_realistic} can be achieved in simulation, under some simplifying assumptions
that we will relax in future work.

\section{Forecasts}
\label{sec:forecasts}

We have now constructed a quadratic estimator $\htau_{\ell m}$ for fluctuations in the optical depth due to patchy reionization,
shown how to compute signal-to-noise (\S\ref{sec:quad_est_realistic}), and tested our signal-to-noise calculation against simulations
(\S\ref{sec:simulation}).  In this section we will make more detailed forecasts.  
All results in this section use the simplified estimator from \S\ref{ssec:simple_estimator}.

\subsection{Signal-to-noise}

\begin{table}
\begin{center}
\begin{tabular}{|c|c|c|c|c|} 
        \hline          &  $\nu$  & $\theta_{\rm FWHM}$ & $\Delta_P$ & $f_{\rm sky}$   \\
\hline  SPTpol  &    $90$ GHz    &  $1.7$ arcmin & $4.3$ $\mu$K-arcmin & $0.015$  \\
&$150$ GHz & $1.0$ arcmin & $4.0$ $\mu$K-arcmin & $0.015$  \\
&$220$ GHz & $0.7$ arcmin & $12.0$ $\mu$K-arcmin & $0.015$  \\
  \hline  EPIC-$2$m  &   $30$ GHz    &  $26.0$ arcmin  & $19.20$ $\mu$K-arcmin & $0.7$  \\ 
&   $45$ GHz    &  $17.0$ arcmin  & $8.27$ $\mu$K-arcmin & $0.7$  \\
&   $70$ GHz    &  $11.0$ arcmin  & $4.19$ $\mu$K-arcmin & $0.7$  \\
&   $100$ GHz    &  $8.0$ arcmin  & $3.24$ $\mu$K-arcmin & $0.7$  \\
&   $150$ GHz    &  $5.0$ arcmin  & $3.13$ $\mu$K-arcmin & $0.7$  \\
&   $220$ GHz    &  $3.5$ arcmin  & $4.79$ $\mu$K-arcmin & $0.7$  \\ 
&  $340$ GHz     &  $2.3$ arcmin  & $21.59$ $\mu$K-arcmin & $0.7$  \\ \hline
\end{tabular}
\end{center}
\caption{Experimental parameters used when making forecasts for SPTpol and EPIC in \S\ref{sec:forecasts}. In each case, we assumed that the lowest and highest frequency channels were used for foreground subtraction, i.e. forecasts were performed using the sensitivity of the 150 GHz channel alone (SPTpol) \cite{SPTpol} or the middle five channels (EPIC), taken from \cite{Baumann:2008aq}.}
\label{tab:sptpol_epic}
\end{table}

\begin{figure}[htbp]
\begin{center}
\includegraphics[width=3.5in]{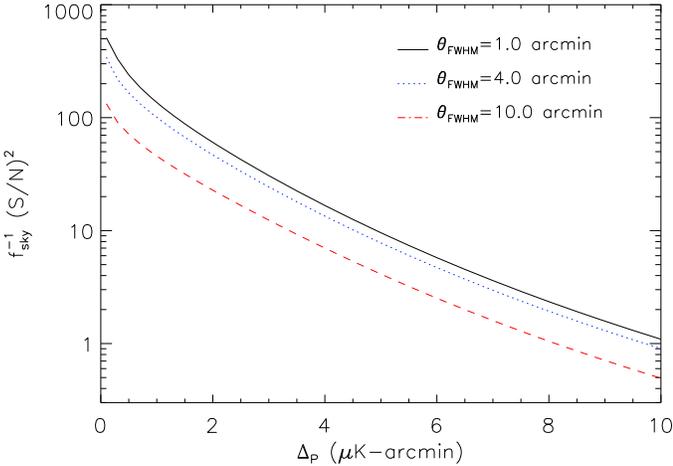}
\end{center}
\caption{Total signal-to-noise of the $\tau$ reconstruction as a function of noise level $\Delta_P$ and beam size $\theta_{\rm FWHM}$, assuming $\ellmax=4000$.}
\label{fig:effective_signal_to_noise}
\end{figure}

In Fig.~\ref{fig:effective_signal_to_noise}, we show the total signal-to-noise of the $\tau$ reconstruction
for a range of noise levels and beam sizes.
A satellite experiment with $\fsky=0.7$ and $\theta_{\rm FWHM}=4$ arcmin can obtain a 3$\sigma$ detection
if the instrumental noise is $\lesssim$ $4$ $\mu$K-arcmin.
For a ground-based experiment with $\fsky=0.05$ and $\theta_{\rm FWHM}=1$ arcmin, the threshold for a $3\sigma$
detection is $\Delta_P=0.7$ $\mu$K-arcmin.
We also forecast total signal-to-noise using parameters for the future SPTpol and EPIC experiments 
(Table ~\ref{tab:sptpol_epic}) and find $(S/N)^2=0.3$ and $(S/N)^2=28$ respectively.
We note that it may be possible to improve these signal-to-noise estimates using delensing or cross-correlating
$\htau_{\ell m}$ with large-scale structure; see \S\ref{sec:discussion} for more discussion.

The preceding forecasts assume that the number of modes that can be measured at a given angular scale is proportional
to $\fsky$.  However, a real experiment using differential detectors on a small patch of sky cannot measure the reionization
E-mode at $\ell\approx 8$, since the wavelength will be larger than the patch size.  For a small-$\fsky$ experiment, the preceding forecasts
will apply without modification if a cosmic variance limited measurement of the large-scale E-mode is available from external
data.  We find that the signal-to-noise is degraded by $\lesssim 40$\% if no external large-scale E-mode measurement is available, or
$\lesssim 10$\% if the large-scale E-mode has been measured with roughly the sensitivity expected for the Planck satellite.

\begin{figure}[htbp]
\begin{center}
\includegraphics[width=3.5in]{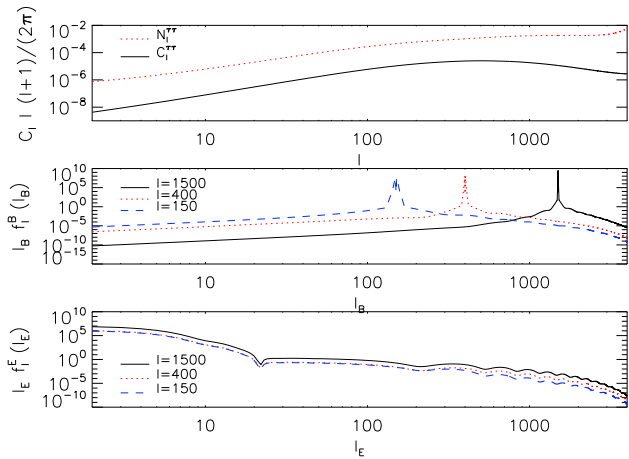}
\end{center}
\caption{{\em Top panel}: Comparison between signal and noise {\em per mode} of the $\Delta\tau$-field, assuming
$\ell_{\rm max}=4000$ and an experiment with $\Delta_P=0.5$ $\mu$K-arcmin and $\theta_{FWHM}=1$ arcmin.
{\em Middle panel:} Distribution of statistical weight $f_\ell^B(\ell_B)$, defined in Eq.~(\ref{eq:flb}), in the B-mode multipole $\ell_B$
for $\tau$ fluctuations reconstructed at $\ell=150,400$ and $1500$, showing a ``spike'' at $\ell_B=\ell$ on top of a broad distribution.
{\em Bottom panel:} Distribution of statistical weight $f_\ell^E(\ell_E)$ in the E-mode multipole $\ell_E$, showing a bump at low $\ell_E$ plus a broad
distribution that extends to high $\ell_E$.}
\label{fig:FlEB_l1l2}
\end{figure}

The total signal-to-noise of the $\tau$ reconstruction can be large enough to give a
many-sigma measurement of the power spectrum, but the signal-to-noise per mode of the $\Delta\tau$-field is always low.
In Fig.~\ref{fig:FlEB_l1l2} (top panel) we compare the signal power spectrum $C_\ell^{\tau\tau}$ to the noise power spectrum $N_\ell^{\tau\tau}$ assuming $\Delta_P=0.5$ $\mu$K-arcmin, $\theta_{FWHM}=1$ arcmin and $f_{\rm sky}=1$. The ratio of the two gives the signal-to-noise per mode, which is $\lesssim 0.025$ for all multipoles, while the overall detection is $\sim 15 \sigma$.

We can ask: how do different angular scales in the CMB contribute to the signal-to-noise of the $\tau$ reconstruction?
To answer this question, we write the inverse noise power spectrum as a sum over ``E multipoles'' $\ell_E$ or over ``B multipoles''
$\ell_B$:
\be
\frac{1}{N_\ell^{\tau\tau}} = \sum_{\ell_E}f_\ell^{E}(\ell_E) = \sum_{\ell_B}f_\ell^{B}(\ell_B)
\ee
\be\label{eq:fle}
f_\ell^{E}(\ell_E)=\frac{1}{2\ell+1} \sum_{\ell'}
\frac{ |\Gamma^{EB(\tau)}_{\ell_E\ell'\ell}|^2 }{ (C_{\ell_E}^{EE} + N_{\ell_E}^{EE}) (C_{\ell'}^{BB} + N_{\ell'}^{BB}) } 
\ee
\be
f_\ell^{B}(\ell_B)=\frac{1}{2\ell+1} \sum_{\ell'}
\frac{ |\Gamma^{EB(\tau)}_{\ell'\ell_B\ell}|^2 }{ (C_{\ell'}^{EE} + N_{\ell'}^{EE}) (C_{\ell_B}^{BB} + N_{\ell_B}^{BB}) } \label{eq:flb}
\ee
When we reconstruct $\tau$ fluctuations on some scale $\ell$, the reconstruction is a quadratic combination
of E-modes and B-modes whose angular scales $\ell_E$ and $\ell_B$ need not be the same as the angular scale $\ell$ of the $\tau$ mode that 
is being estimated.
The quantities $f_\ell^E(\ell_E)$ and $f_\ell^B(\ell_B)$ defined in Eqs.~(\ref{eq:fle}) and (\ref{eq:flb})
can be interpreted as the distribution of statistical weight in $\ell_E$ and $\ell_B$ for
a fixed $\ell$.

In Fig.~\ref{fig:FlEB_l1l2} (middle and lower panels), we show $f_\ell^B(\ell_B)$ and $f_\ell^E(\ell_E)$ for three different values of $\ell$.
It is seen that the statistical weight is a sum of two contributions.
The first contribution appears as a ``bump'' with $\ell_E \lesssim 30$ and a ``spike'' at $\ell_B\approx \ell$.
This three-way correlation between $(E,B,\tau)$ can be interpreted as modulation of the large-scale reionization 
E-mode by a smaller-scale $\tau$ fluctuation,
generating a B-mode with nearly the same wavelength as the $\tau$ fluctuation.
(Note that if $\ell_E$ is small, then $\ell_B$ must be $\approx\ell$ because the triple $(\ell_E,\ell_B,\ell)$
must satisfy the triangle inequality.)
The second contribution comes from a wide range of multipoles $\ell_E$ and $\ell_B$ with a shape which is roughly
independent of $\ell$.
This contribution can be interpreted as arising from the $e^{-\tau}$ screening of the acoustic peaks, which generates
B-modes across a wide range of angular scales.

These two regimes can be considered separately, and one can construct two quadratic estimators $\htau_{\ell m}$: one coming from a region with $\ell_E \le 30$ and the other one coming from $\ell_E \ge 30$. The two estimators could be cross correlated in order to have a more robust power spectrum estimate, or they could be subtracted from each other, as a null test. We find that the detection significance obtained using this cross-correlation is $\approx 70$\% (in the limit of low instrumental sensitivity) to 
$\approx 40$\% (in the limit of high sensitivity) of the total detection significance.

\subsection{Reionization model parameters}
\label{ssec:fisher}

What can we learn about patchy reionization by measuring the $C_\ell^{\tau\tau}$ power spectrum?
As a first look at this question, we will forecast parameter uncertainties in the five-parameter model
$\{\bar R, \sigma_{\ln R}, \tau, \Delta_y, b\}$ described in \S\ref{sec:reionization_model}.

This simple semianalytic model
does not contain some of the qualitative features seen in simulations, 
such as the dependence of the characteristic size of the ionized regions on the redshift \cite{Zahn:2006sg}, 
or the dependence of the bubble bias on the size of the ionized bubbles \cite{McQuinn:2005ce}.
However, we will restrict our scope to the five-parameter model to get a qualitative sense of the information contained in $C_\ell^{\tau\tau}$.
All forecasts in this section are for the $EB$ estimator, assuming $\Delta_P=0.5$ $\mu$K-arcmin, $\theta_{\rm FWHM}=1$ arcmin and $f_{\rm sky}=0.7$.

Our tool for forecasting will be the Fisher matrix: if the power spectrum $C_\ell^{\tau\tau}$ is a function
of $N$ parameters $\{\pi_1,\ldots,\pi_N\}$, then we compute the $N$-by-$N$ matrix as
\be
F_{ij} = \frac{\fsky}{2} \sum_\ell (2\ell+1) \frac{(\partial C_\ell^{\tau\tau}/\partial\pi_i) (\partial C_\ell^{\tau\tau}/\partial\pi_j)}{(C_\ell^{\tau\tau}+N_\ell^{\tau\tau})^2}
\ee
The rms uncertainty on parameter $\pi_i$ if the remaining $(N-1)$ parameters are assumed fixed is given by $\sigma(\pi_i) = (F_{ii})^{-1/2}$; 
if the remaining parameters are assumed marginalized then the uncertainty is $\sigma(\pi_i) = (F^{-1}_{ii})^{1/2}$.

Some caveats should be kept in mind when interpreting Fisher matrix forecasts.
The Fisher forecasting procedure can be interpreted as extrapolating the likelihood function ${\mathcal L}[\pi_i|{\rm data}]$
to the entire parameter space by computing second derivatives at maximum likelihood, and assuming that the likelihood
function is a multivariate Gaussian.
This should be a good approximation in directions of parameter space which correspond to well-constrained combinations of parameters,
but in poorly-constrained directions the approximation may not be accurate.
For example, in a parameter space which contains degeneracies, the $1\sigma$ region of parameter space can have a complex shape but the
Fisher forecast will always approximate it by an ellipse-shaped region oriented in a direction which is tangent to the true region.
In this situation, the Fisher matrix is still useful for identifying
the presence of a degeneracy, and the corresponding direction in parameter space, but the forecasted uncertainty in the 
degenerate direction is not necessarily accurate.

\begin{figure}[htbp]
\begin{center}
\includegraphics[width=3.3in]{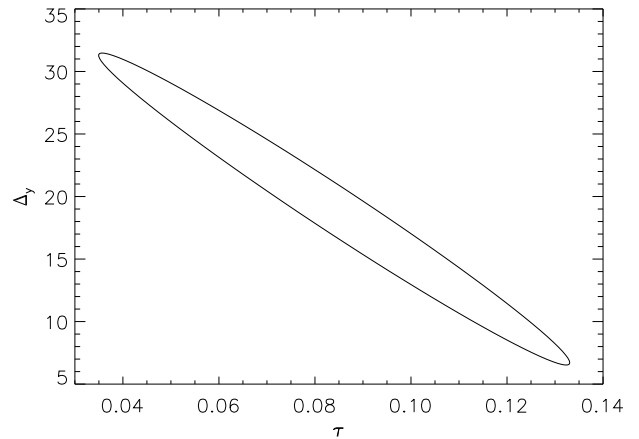}
\end{center}
\caption{Forecasted uncertainties on the total optical depth $\tau$ and reionization width parameter $\Delta_y$,
assuming that the remaining parameters of the model $\{\bar R, \sigma_{\ln R}, b\}$ are fixed to fiducial values.
Ellipses here and throughout \S\ref{ssec:fisher} are plotted at $\Delta\chi^2=1$ and not 68\% C.L.}
\label{fig:error_ellipse_tau_Deltay}
\end{figure}

\begin{figure}[htbp]
\begin{center}
\includegraphics[width=3.0in]{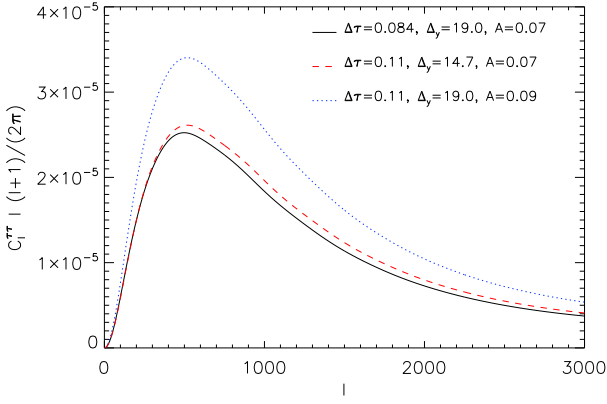}
\end{center}
\caption{Power spectrum $C_\ell^{\tau\tau}$ for different values of the optical depth $\tau$
and reionization width parameter $\Delta_y$.
Changing either $\tau$ or $\Delta_y$ produces a change in the overall amplitude of the power spectrum, but if these parameters are varied jointly in a way
which preserves the value of $A$ (defined in Eq.~(\ref{eq:a_def})), then the amplitude remains approximately fixed.}
\label{fig:fisher_plot1}
\end{figure}

As a first step toward understanding the five-parameter space, consider the forecasted uncertainties on the parameters $\{ \tau, \Delta_y \}$
with the remaining model parameters $\{b,\bar R,\sigma_{\ln R}\}$ fixed to fiducial values.
Each of $\tau$ and $\Delta_y$ can be constrained to $\approx 11\%$ if the other one is assumed fixed.
The error ellipse in the Fisher approximation is shown in Fig.~\ref{fig:error_ellipse_tau_Deltay}.
It is seen that the parameters $\tau$ and $\Delta_y$ are nearly degenerate (the correlation coefficient is $-0.985$).
The interpretation is that only one combination of $\tau$ and $\Delta_y$ is observable.
To a good approximation (see Fig.~\ref{fig:fisher_plot1}), the observable parameter is simply the overall amplitude $A$ of the $\tau$-power spectrum,
which we define by:
\be
A = \int \frac{dz}{H(z)} \frac{(1+z)^4}{\chi(z)^2} x_e(z) [ 1 - x_e(z) ]  \label{eq:a_def}
\ee
To motivate this definition of $A$, note that if $P_{\Delta x_e \Delta x_e}$ is a slowly varying function of $\chi$ and $k$ in Eq.~(\ref{eq:cltau}),
then $C_\ell^{\tau\tau}$ will depend on the reionization history only through an overall factor of $A$.
(As a technical point,
note that this integral formally diverges as $z\rightarrow 0$, because $x_e$ does not precisely equal $1$ in our fiducial model,
but $\chi(z)\rightarrow 0$ in the denominator.  We took $\zmin=0.1$ and tested that the value of the integral is stable over the range [0.01,1.0].)

\begin{figure}[htbp]
\begin{center}
\includegraphics[width=3.5in]{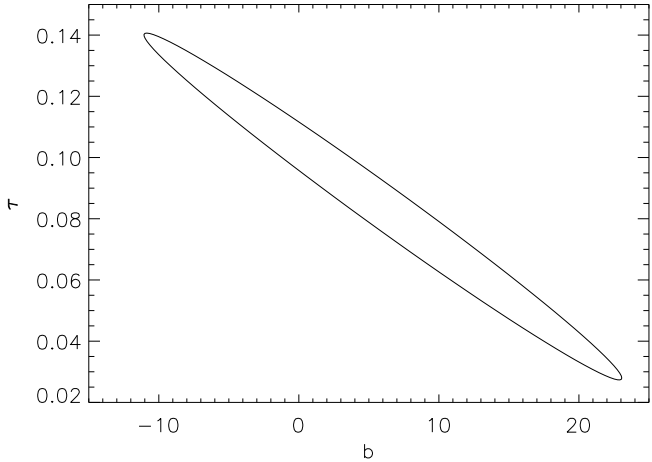}
\end{center}
\caption{Error ellipse in the Fisher approximation for the bubble bias $b$ and total optical depth $\tau$, 
with the remaining parameters of the model $\{ \bar R, \sigma_{\ln R}, \Delta_y \}$ assumed fixed to fiducial values.}
\label{fig:error_ellipse_bias_tau}
\end{figure}

\begin{figure}[htbp]
\begin{center}
\includegraphics[width=3.0in]{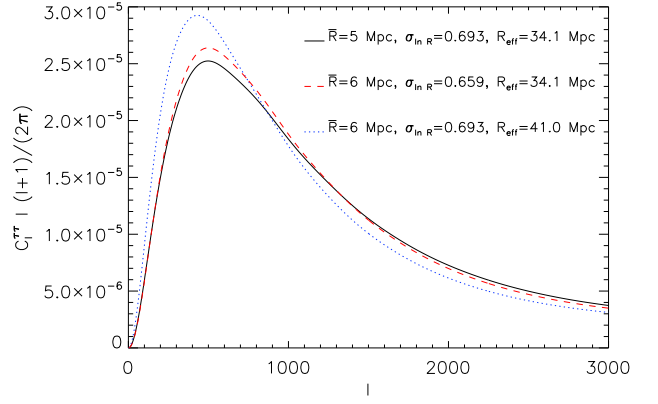}
\end{center}
\caption{Power spectrum $C_\ell^{\tau\tau}$ for different values of the mean $\bar R$ and width $\sigma_{\ln R}$ of the log-normal bubble radius distribution.
Changing either $\bar R$ or $\sigma_{\ln R}$ will change the location in $\ell$ where the $\tau$-power spectrum peaks, but if the two parameters 
are jointly varied to preserve the value of $R_{\rm eff}$ (defined in Eq.~(\ref{eq:reff_def})), then the peak location is approximately unchanged.}
\label{fig:fisher_plot2}
\end{figure}

Considering next the bubble bias $b$, we find that $b$ and $\tau$ are nearly 100\% correlated (Fig.~\ref{fig:error_ellipse_bias_tau}).
The bias can be thought of as a third parameter (together with $\tau$ and $\Delta_y$) which contributes to the overall amplitude of the power spectrum;
however the dependence of the amplitude on $b$ is relatively weak: a change of $\Delta b=1$ produces a $\approx 3$\% change in $C_\ell^{\tau\tau}$.

\begin{figure}[htbp]
\begin{center}
\includegraphics[width=3.3in]{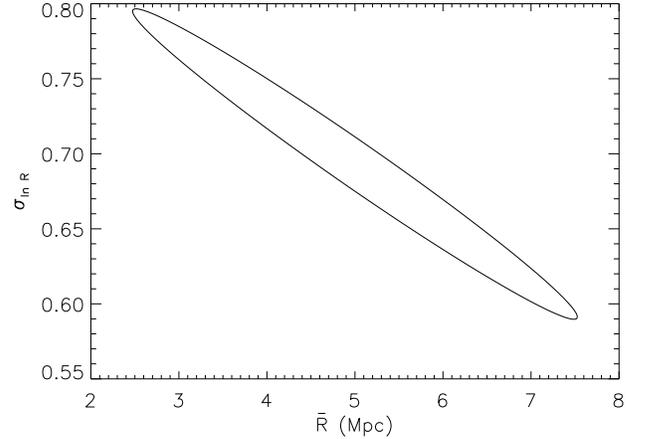}
\end{center}
\caption{Error ellipse for the characteristic bubble size $\bar R$ and dispersion $\sigma_{\ln R}$,
assuming that the remaining parameters of the model $\{ \tau, \Delta_y, b \}$ are fixed to their fiducial values.}
\label{fig:error_ellipse_Reff_Rbar}
\end{figure}

Now consider the parameters $\{\bar R, \sigma_{\ln R}\}$.
In Fig.~\ref{fig:error_ellipse_Reff_Rbar}, we show the error ellipse in the $\{ \bar R, \sigma_{\ln R} \}$ plane, assuming that the remaining 
parameters $\{ \tau, \Delta_y, b \}$ are fixed to their fiducial values.
The parameters $\bar R$ and $\sigma_{\ln R}$ are degenerate but each can be constrained to $\approx 8$\% if the other is assumed fixed.
This is completely analogous to the $(\tau, \Delta_y)$ degeneracy described previously: there is only one observable combination of the two parameters.
Here, we interpret the observable as the peak location in $\ell$, which we define by:
\be
R_{\rm eff}= \bar{R}e^{4\sigma^2_{\ln R}}  \label{eq:reff_def}
\ee
This definition follows \cite{Mortonson:2006re}: $R_{\rm eff}$ is the scale where the square of the averaged window function $\langle W(kR)^2\rangle$ makes a transition from low to high values of $k$.
The peak location degeneracy is illustrated directly in Fig.~\ref{fig:fisher_plot2}.

In conclusion, the Fisher matrix analysis in this section has shown that in the five-parameter space $\{ \bar R, \sigma_{\ln R}, \tau, \Delta_y, b \}$,
there are only two independent observables: first, an ``amplitude'' observable $A$ which is as a function of $\{ \tau, \Delta_y, b \}$ and second,
a ``peak location'' observable $R_{\rm eff}$ which is a function of $\{ \bar R, \sigma_{\ln R} \}$.
The two observables are correlated but not degenerate: we find a correlation coefficient $\approx -0.5$ between $A$ and $R_{\rm eff}$.

The amplitude $A$ of the $\tau$-power spectrum only depends on the mean ionization fraction $\bar x_e(z)$.
The mean reionization history can also be constrained using the large-scale E-mode, but 
$A$ isolates the patchy epoch in the sense that it only acquires contributions from redshifts for which reionization is inhomogeneous ($0 < x_e(z) < 1$).
A comparison between parameter constraints from the $\tau$-power spectrum and the large-scale E-mode is likely to depend strongly on the parameterization
of the reionization history which is assumed.

Another potential source of information is the small-scale temperature power spectrum \cite{Zhang:2003nr},
which is sensitive to patchy reionization via the kSZ effect, but
requires careful modeling of other secondaries on small scales (e.g. gravitational lensing, the kSZ effect from low-redshift large-scale structure,
thermal SZ, and the Rees-Sciama effect).
It would be interesting to compare parameter constraints from the $\tau$-power spectrum and the small-scale temperature power spectrum,
but this is outside the scope of a first paper.
One advantage of working with the $\tau$ reconstruction is that the 
greatest signal-to-noise comes from modes on sufficiently large scales ($\ell\sim 400$)
that the patchy reionization contribution dominates the low-redshift contribution from large-scale structure.
(This can be seen directly by comparing the two curves in Fig.~\ref{fig:xe2}.)
The $\tau$-power spectrum is also weighted toward high redshifts (relative to the small-scale temperature)
because the kSZ power spectrum contains a factor $\langle v^2 \rangle$ and the velocity power spectrum is growing in time.

\section{Discussion}
\label{sec:discussion}

In this paper we have introduced a new technique for extracting the patchy reionization signal from the cosmic microwave background.
In a fixed realization of the $\Delta\tau$-field, the two-point function of the CMB acquires terms which are linear in $\Delta\tau$, and
this allows us to construct a quadratic estimator $\htau_{\ell m}$ for the modes of the $\Delta\tau$-field.
This construction is formally similar to the lens reconstruction estimator $\hphi_{\ell m}$ which estimates modes of the CMB lens potential $\phi$ using the induced two-point function of the observed CMB.

In principle the general construction in this paper can use all combinations of the \{T,E,B\} fields,
and can distinguish transverse modes of the $\Delta\tau$-field at different 
redshifts using the redshift dependence of the induced CMB two-point function.
However, we find that in practice, the redshift dependence is so weak that only one 2D field $(\Delta\tau)_{\ell m}$ can be reconstructed, and that
nearly all the signal-to-noise comes from the $EB$ quadratic estimator (with contributions from Thomson and screening effects which are comparable
in signal-to-noise).
This analysis also implies that it is possible to write the estimator in a simple form (\S\ref{ssec:simple_estimator}) which ignores subtleties like the redshift dependence.

It is worth emphasizing that the quadratic estimator framework depends on having a large cross-correlation between the new CMB anisotropy generated by patchy
reionization and the primary CMB.
This is why our estimator is sensitive to the Thomson and screening effects during patchy reionization: in both of these cases, the new CMB anisotropy
is highly correlated to the large-scale reionization E-mode, or to the acoustic E-mode, respectively.
However, the quadratic estimator is only sensitive to the kSZ effect in temperature via its cross-correlation to the Doppler effect on large scales.
Another way of saying this is that the quadratic estimator can only ``see'' the (Doppler, kSZ) cross-correlation but not the (kSZ, kSZ) power spectrum,
since the two are $\bigoh(\Delta\tau)$ and $\bigoh(\Delta\tau^2)$ effects respectively.
For this reason, we consider it a separate unsolved problem, not addressed in this paper, to construct a statistic which can separate the kSZ signal in temperature
from other sources of small-scale power such as CMB lensing or the primary anisotropy.

In this paper we have done a complete analysis of the signal-to-noise that can be achieved using a quadratic estimator construction,
concluding with a Fisher matrix forecast for a simple five-parameter reionization model (\S\ref{ssec:fisher}).  In this model, there are
parameter degeneracies that leave only two well-constrained power spectrum observables: an ``amplitude'' observable $A$ which is mainly
sensitive to the duration of patchy reionization, and a ``peak location'' observable $R_{\rm eff}$ which is mainly sensitive to the size distribution of bubbles.

Let us conclude by commenting on some possible continuations of this work which are outside the scope of a first paper.

One such extension is to increase the realism of the reionization model, and to study prospects for constraining a more complex parameter set.
It is unclear whether the ``two-observable'' picture that we obtained in our Fisher matrix analysis of the five-parameter model will be qualitatively
different in a more complete model.

Another natural continuation is to study prospects for cross-correlation with other flavors of cosmological data which probe the epoch of
inhomogeneous reionization.
A nice feature of the quadratic estimator framework (as compared to other statistics such as the CMB power spectrum) is that the patchy reionization 
signal is extracted in the form of a map $\htau_{\ell m}$ which can be cross-correlated to other datasets.
This cross-correlation technique is part of future work, and can be used either to boost signal-to-noise for a first detection (e.g. \cite{Smith:2007rg,Hirata:2008cb}
in the context of CMB lens reconstruction) or to make the measurement more robust by reducing the range of systematic effects which may contaminate 
the auto power spectrum of the reconstruction.

Finally, a continuation of this work which is currently in preparation
is the implementation of the estimator in more realistic simulations which include
the non-Gaussian statistics of the lensed B-mode.
The simulations in this paper (\S\ref{sec:simulation}) show that our signal-to-noise forecasts can be achieved in toy Monte Carlo simulations
in which the lensed B-mode has the correct power spectrum but the statistics are treated as Gaussian.
Including non-Gaussianity is an important step: it can potentially bias the estimator $\htau_{\ell m}$ (which reconstructs the modes of the 
$\Delta\tau$-field by using the induced non-Gaussian signature) but also presents the opportunity for improving signal-to-noise by combining $\tau$ reconstruction with delensing.

\section*{Acknowledgements}

We would like to thank Anthony Challinor, Chris Hirata, Antony Lewis, Adam Lidz, Michael Mortonson, David Spergel, Bruce Winstein, and Oliver Zahn for useful discussions;
and Wayne Hu for frequent discussions and feedback throughout the project.
This work was supported by the KICP through the Grant No. NSF PHY-0114422 (CD) and by the STFC (KMS).
KMS would like to thank the hospitality of the Department of Astrophysics at Princeton University, where this work was partially carried out.

\vfill
\bibliography{paper1v2}

\appendix

\section{Quadratic estimator formalism}
\label{sec:optimal_estimator}

The purpose of this appendix is to develop the quadratic estimator construction that was used in \S\ref{sec:simple_estimator},~\S\ref{sec:quad_est_realistic}
and to establish some key properties:
the form of the quadratic estimator (Eqs.~(\ref{eq:quadratic_estimator}),~(\ref{eq:eigenmode_htau})),
the noise covariance (Eqs.~(\ref{eq:sec4_nl}),~(\ref{eq:eigenmode_nl})),
and the total signal-to-noise (Eq.~(\ref{eq:eigenmode_total_sn})).
In \S\ref{ssec:single_field} we will develop the ``single-field'' version that was used assuming the constant quadrupole approximation in \S\ref{sec:simple_estimator}.
In \S\ref{ssec:multiple_fields} we will generalize to the multifield case that was used in \S\ref{sec:quad_est_realistic} to correctly treat the redshift dependence.

\subsection{Quadratic estimator formalism: single field}
\label{ssec:single_field}

Let us first make the approximation that patchy reionization is narrow, so that the anisotropic optical depth can be described by a single 2D field $(\Delta\tau)_{\ell m}$.
In \S\ref{sec:quad_est_realistic}, we showed that to lowest order in $\Delta\tau$, patchy reionization produces correlations between CMB multipoles $a_{\ell m} \ne a_{\ell'm'}$ which
are of the form:
\be
\left\langle a_{\ell_1 m_1}^{X}a_{\ell_2m_2}^{Y}\right\rangle=\sum_{\ell m}\Gamma^{XY(\tau)}_{\ell_1\ell_2\ell}\wj{\ell_1}{\ell_2}{\ell}{m_1}{m_2}{m} (\Delta\tau)^{*}_{\ell m}\,,  \label{eq:gamma_2pt}
\ee
where $X,Y\in\{T,E,B\}$, and the couplings $\Gamma^{XY(\tau)}_{\ell_1\ell_2\ell}$ are given in Eqs.~(\ref{eq:gamma_first})--(\ref{eq:gamma_last}).

Given the two-point function in Eq.~(\ref{eq:gamma_2pt}), we construct a quadratic estimator $\htau_{\ell m}$ for each mode of the $\Delta\tau$-field,
by solving for the weights $W^{XY}_{\ell m\ell_1 m_1\ell_2 m_2}$ in
\be
\htau^{*}_{\ell m} = \frac{1}{2} \sum_{XY}\sum_{\ell_1m_1\ell_2m_2} W^{XY}_{\ell m\ell_1 m_1\ell_2 m_2} a^X_{\ell_1m_1} a^Y_{\ell_2m_2},
\ee
which minimize the variance
\ba
\Var(\htau^{*}_{\ell m}) &=& \frac{1}{2} \sum_{XYX'Y'}\sum_{\ell_1m_1\ell_2m_2} W^{XY}_{\ell m\ell_1m_1\ell_2m_2} \nn \\
&& \hskip 0.2in \times W^{X'Y'*}_{\ell'm'\ell_1m_1\ell_2m_2} C^{XX'}_{\ell_1} C^{YY'}_{\ell_2},
\ea
subject to the normalization constraint $\langle \htau_{\ell m} \rangle = (\Delta\tau)_{\ell m}$:
\be
\frac{1}{2} \sum_{\ell_1m_1\ell_2} W^{XY}_{\ell m\ell_1m_1\ell_2m_2} \Gamma^{XY(\tau)}_{\ell_1\ell_2\ell} \wj{\ell_1}{\ell_2}{\ell}{m_1}{m_2}{m} = 1  \label{eq:single_field_normalization}
\ee
A short calculation shows that the minimum-variance estimator is given by:
\ba
\htau_{\ell m} &=& {N_\ell^{\tau\tau}\over2}
\sum_{XYX'Y'\ell_1m_1\ell_2m_2}
\Gamma^{XY(\tau)}_{\ell_1\ell_2\ell}
\wj{\ell_1}{\ell_2}{\ell}{m_1}{m_2}{m}  \nn \\
&&
\hskip 0.2in \times
({\bf C}^{-1})_{\ell_1}^{XX'}a^{X'*}_{\ell_1m_1} ({\bf C}^{-1})_{\ell_2}^{YY'}a^{Y'*}_{\ell_2m_2},  \label{eq:min_var_single_field}
\ea
where:
\ba
\frac{1}{N_{\ell}^{\tau\tau}} &=& \frac{1}{2(2\ell+1)} \sum_{XYX'Y'\ell_1\ell_2}
\Gamma^{XY(\tau)}_{\ell_1\ell_2\ell}
({\bf C}^{-1})^{XX'}_{\ell_1} \nn \\
&& \hskip 0.2in \times
\Gamma^{X'Y'(\tau)*}_{\ell_1\ell_2\ell}
({\bf C}^{-1})^{YY'}_{\ell_2}  \label{eq:fl_def},
\ea
where ${\bf C}_\ell$ is defined in Eq. (\ref{eq:Cmatrix_def}).
Now that we have found the optimal estimator, we consider its signal-to-noise.
Let us carefully distinguish between what we mean by ``signal'' and what we mean by ``noise'', in the context of $\tau$ reconstruction.

We write the two-point function of $\htau_{\ell m}$ as the sum of two terms:
\ba
\left\langle \htau^*_{\ell m} \htau_{\ell'm'} \right\rangle &=& 
   \left\langle \htau^*_{\ell m} \htau_{\ell'm'} \right\rangle_{\rm noise} +
   \left\langle \htau^*_{\ell m} \htau_{\ell'm'} \right\rangle_{\rm signal}  \nn \\
\left\langle \htau^*_{\ell m} \htau_{\ell'm'} \right\rangle_{\rm noise} &=& N_\ell^{\tau\tau} \delta_{\ell\ell'} \delta_{mm'} \nn \\
\left\langle \htau^*_{\ell m} \htau_{\ell'm'} \right\rangle_{\rm signal} &=& C_\ell^{\tau\tau} \delta_{\ell\ell'} \delta_{mm'} \label{eq:signal_noise_split}
\ea
The first term $\langle \htau^*_{\ell m} \htau_{\ell'm'} \rangle_{\rm noise}$ is obtained by summing Gaussian contractions between CMB multipoles 
$a_{\ell m}^X$ and is what we mean by ``noise'': it is the power spectrum of
the reconstruction due solely to statistical fluctuations in the CMB, in the absence of any patchy reionization signal.
The second term $\langle \htau^*_{\ell m} \htau_{\ell'm'}\rangle_{\rm signal}$ arises from the expectation value $\langle\htau_{\ell m}\rangle = (\Delta\tau)_{\ell m}$
and is what we mean by ``signal'': excess power in the reconstruction due to the presence of $\tau$ fluctuations.
We therefore interpret $N_\ell^{\tau\tau}$ as the reconstruction noise power spectrum and $C_\ell^{\tau\tau}$ as the signal
power spectrum.

In a patchy reionization model with signal power spectrum $C_\ell^{\tau\tau}$, the total signal-to-noise of the reconstruction (summed over all modes) is
\be
S/N = \left[ \frac{\fsky}{2} \sum_\ell (2\ell+1) \left( \frac{C_\ell^{\tau\tau}}{N_\ell^{\tau\tau}} \right)^2 \right]^{1/2} \label{eq:total_sn_single_field}
\ee
This is the ``number of sigmas'' for an overall detection of patchy reionization, via excess power in the $\tau$ reconstruction.

Our construction of $\htau_{\ell m}$ is formally identical to the quadratic estimator $\hphi_{\ell m}$ for the CMB lens potential $\phi$.
More precisely, to lowest order in $\phi$, the two-point function induced by lensing is of the form of Eq.~(\ref{eq:gamma_2pt}), with the $\Delta\tau$-field
replaced by $\phi$, and the object $\Gamma^{XY(\tau)}_{\ell_1\ell_2\ell}$ replaced by
\ba
\Gamma^{TT(\phi)}_{\ell_1\ell_2\ell_3} &=& C_{\ell_1}^{TT} F^0_{\ell_2\ell_1\ell_3} + C_{\ell_2}^{TT} F^0_{\ell_1\ell_2\ell_3} \nn \\
\Gamma^{TE(\phi)}_{\ell_1\ell_2\ell_3} &=& C_{\ell_1}^{TE} \left( \frac{F^{-2}_{\ell_2\ell_1\ell_3} + F^{2}_{\ell_2\ell_1\ell_3}}{2} \right) + C_{\ell_2}^{TE} F^0_{\ell_1\ell_2\ell_3} \nn \\
\Gamma^{EE(\phi)}_{\ell_1\ell_2\ell_3} &=& C_{\ell_1}^{EE} \left( \frac{F^{-2}_{\ell_2\ell_1\ell_3} + F^{2}_{\ell_2\ell_1\ell_3}}{2} \right) \nn \\
                           && + C_{\ell_2}^{EE} \left( \frac{F^{-2}_{\ell_1\ell_2\ell_3} + F^{2}_{\ell_1\ell_2\ell_3}}{2} \right)   \nn \\
\Gamma^{TB(\phi)}_{\ell_1\ell_2\ell_3} &=& C_{\ell_1}^{TE} \left( \frac{F^{-2}_{\ell_2\ell_1\ell_3} - F^{2}_{\ell_2\ell_1\ell_3}}{2i} \right) \nn  \\
\Gamma^{EB(\phi)}_{\ell_1\ell_2\ell_3} &=& C_{\ell_1}^{EE} \left( \frac{F^{-2}_{\ell_2\ell_1\ell_3} - F^{2}_{\ell_2\ell_1\ell_3}}{2i} \right), \label{eq:gamma_lensing}
\ea
where the $F$ symbol is defined by:
\ba
F^s_{\ell_1\ell_2\ell_3} = [-\ell_1(\ell_1+1) + \ell_2(\ell_2+1) + \ell_3(\ell_3+1)] \nn\\
\times\sqrt{\frac{(2\ell_1+1)(2\ell_2+1)(2\ell_3+1)}{16\pi}} \wj{\ell_1}{\ell_2}{\ell_3}{-s}{s}{0}
\ea

With these replacements, it can be checked that the estimator we have derived in Eq.~(\ref{eq:min_var_single_field}) agrees with the well-known minimum-variance 
quadratic estimator $\hphi_{\ell m}$ for lens reconstruction.
In the single-field case, the quadratic estimators for lens reconstruction and patchy reionization can simply be regarded as arising from a different
set of couplings $\Gamma^{XY}_{\ell_1\ell_2\ell}$.
(See also \cite{Kamionkowski:2008fp} for another example of the quadratic estimator formalism: reconstructing a spatially fluctuating rotation $\alpha(\n)$ in the CMB polarization.)

\subsection{Quadratic estimator formalism: multiple fields}
\label{ssec:multiple_fields}

The quadratic estimator formalism in the preceding subsection can be applied whenever the CMB two-point function $\langle a_{\ell m}^X a_{\ell' m'}^Y \rangle$ is 
proportional to an auxiliary field which is not directly observable and must be estimated, such as $\Delta\tau$ or $\phi$.
(The form of the two-point function given in Eq.~(\ref{eq:gamma_2pt}) is the most general form allowed by global rotation invariance.)

However, in \S\ref{sec:quad_est_realistic} we will need greater generality:
we are also interested in the case where there are $N$ auxiliary fields $\Delta\tau^{\alpha}$, where $\alpha=1,\ldots,N$
runs over redshift bins.

We consider a CMB two-point function of the general form
\be
\left\langle a_{\ell_1 m_1}^{X}a_{\ell_2m_2}^{Y}\right\rangle=\sum_{\ell m\alpha}\Gamma^{XY(\tau_\alpha)}_{\ell_1\ell_2\ell}\wj{\ell_1}{\ell_2}{\ell}{m_1}{m_2}{m} (\Delta\tau^{\alpha})^{*}_{\ell m},
\ee
for some set of couplings $\Gamma^{XY(\tau_\alpha)}_{\ell_1\ell_2\ell}$.
This form of the two-point function keeps track of the (weak) dependence on the redshift bin.

We propose two possibilities for constructing a quadratic estimator in the multifield case.
The first possibility (which corresponds to the ``simple estimator'' in \S\ref{ssec:simple_estimator})
is simply to use the single-field quadratic estimator corresponding to a fixed redshift bin $\mu$:
\ba
\htau^{(\mu)}_{\ell m} &=& {N_\ell^{\tau\tau (\mu)}\over2}
\sum_{XYX'Y'\ell_1m_1\ell_2m_2}
\Gamma^{XY(\tau_\mu)}_{\ell_1\ell_2\ell}
\wj{\ell_1}{\ell_2}{\ell}{m_1}{m_2}{m}  \nn \\
&&
\hskip 0.2in \times
({\bf C}^{-1})_{\ell_1}^{XX'}a^{X'*}_{\ell_1m_1} ({\bf C}^{-1})_{\ell_2}^{YY'}a^{Y'*}_{\ell_2m_2} \\
\frac{1}{N_{\ell}^{\tau\tau (\mu)}} &=& \frac{1}{2(2\ell+1)} \sum_{XYX'Y'\ell_1\ell_2}
\Gamma^{XY(\tau_\mu)}_{\ell_1\ell_2\ell}
({\bf C}^{-1})^{XX'}_{\ell_1} \nn \\
&& \hskip 0.2in \times
\Gamma^{X'Y'(\tau_\mu)*}_{\ell_1\ell_2\ell}
({\bf C}^{-1})^{YY'}_{\ell_2} 
\ea
The only new ingredient in the multifield case is the nontrivial redshift response: the expectation value is given by
\be
\langle \htau^{(\mu)}_{\ell m} \rangle = \sum_\alpha (R_\ell^{(\mu)}(z^\alpha)) (\Delta\tau^\alpha)_{\ell m},
\ee
where
\ba
R_\ell^{(\mu)}(z^\alpha) &=& 
 \frac{N_\ell^{\tau\tau (\mu)}}{2(2\ell+1)} \sum_{XYX'Y'\ell_1\ell_2}
\Gamma^{XY(\tau_\mu)}_{\ell_1\ell_2\ell}
({\bf C}^{-1})^{XX'}_{\ell_1} \nn \\
&& \hskip 0.2in \times
\Gamma^{X'Y'(\tau_\alpha)*}_{\ell_1\ell_2\ell}
({\bf C}^{-1})^{YY'}_{\ell_2} 
\ea
(Note that $R_\ell^{(\mu)}(z^\mu)=1$.)

It follows that the signal power spectrum of the ``simple'' quadratic estimator is given by
\be
\langle \htau_{\ell m}^{(\mu)*} \htau_{\ell' m'}^{(\mu)} \rangle_{\rm signal} = \sum_\alpha(R_\ell^{(\mu)}(z^\alpha))^2 C_\ell^{\tau_\alpha \tau_\alpha}\delta_{\ell\ell'}\delta_{mm'}
\ee
and the total signal-to-noise is given by:
\be
S/N = \left[\frac{\fsky}{2} \sum_\ell (2\ell+1) \left( 
\frac{\sum_\alpha (R_\ell^{(\mu)}(z^\alpha))^2 C_\ell^{\tau_\alpha \tau_\alpha}}{N_\ell^{\tau\tau (\mu)}}
\right)^2\right]^{1/2}  \label{eq:sn2_simple}
\ee
This construction has the advantage of preserving the simplicity of the single-field estimator in the multifield case,
but the disadvantage that it does not necessarily extract all the signal-to-noise.

The rest of this appendix is devoted to a more complicated eigenmode construction 
(this corresponds to the principal component analysis in \S\ref{ssec:pca})
which is mathematically guaranteed to contain all the signal-to-noise.
For an arbitrary set of couplings $\Gamma^{XY(\tau_\alpha)}_{\ell_1\ell_2\ell}$, the total $(S/N)$ of the simple estimator (Eq.~(\ref{eq:sn2_simple}))
will be less than the total $(S/N)$ of the eigenmode estimator (Eq.~(\ref{eq:sn2_eigenmode_sum})).
However, in our fiducial reionization model, the calculations in \S\ref{ssec:simple_estimator} show that
the signal-to-noise is almost the same in the two cases.
Therefore, in practice the eigenmode construction will serve as a proof that the simple construction does not lose information.

For each $\ell$ we define an $N$-by-$N$ matrix
\ba
F^{\alpha\beta}_\ell &=&  \frac{1}{2(2\ell+1)} \sum_{XYX'Y'\ell_1\ell_2}
\Gamma^{XY(\tau_\alpha)}_{\ell_1\ell_2\ell}
({\bf C}^{-1})^{XX'}_{\ell_1} \nn \\
&& \hskip 0.2in \times
\Gamma^{X'Y'(\tau_\beta)*}_{\ell_1\ell_2\ell}
({\bf C}^{-1})^{YY'}_{\ell_2}  \label{eq:f_matrix_def}
\ea

For each $\ell$, we define weights $w^{(i)}_\ell(z^\alpha)$ by solving the eigenmode equation:
\be\label{eq:lambda_def}
\sum_\beta C_\ell^{\tau_\alpha\tau_\alpha} F_\ell^{\alpha\beta}w^{(i)}_\ell(z^\beta) = \lambda^{(i)}_\ell w^{(i)}_\ell(z^\alpha),
\ee
where the eigenvalues $\lambda^{(i)}_\ell$ satisfy $\lambda^{(1)}_\ell > \lambda^{(2)}_\ell > \cdots > \lambda^{(N)}_\ell$.

We then define an eigenmode estimator by:
\ba
\hE^{(i)}_{\ell m} =
{N_\ell^{\tau\tau (i)}\over2} \sum_\alpha w^{(i)}_\ell(z^\alpha)
\sum_{\ell_1\ell_2}^{\ell_{\rm max}}\sum_{XYX'Y'm_1m_2}
\Gamma^{XY(\tau_\alpha)}_{\ell_1\ell_2\ell}  \nn \\
\times
\wj{\ell_1}{\ell_2}{\ell}{m_1}{m_2}{m}
({\bf C}^{-1})^{XX'}_{\ell_1}a^{X'*}_{\ell_1m_1} ({\bf C}^{-1})^{YY'}_{\ell_2}a^{Y'*}_{\ell_2m_2}
\ea
\be
\frac{1}{N_\ell^{\tau\tau (i)}} =
\sum_{\alpha\beta} w^{(i)}_\ell(z^\alpha) w^{(i)}_\ell(z^\beta) F_\ell^{\alpha\beta}
\ee
The mean response of the eigenmode estimator is given by:
\ba
\langle \hE^{(i)}_{\ell m} \rangle &=& \sum_\alpha R^{(i)}_\ell(z^\alpha) (\Delta\tau^\alpha)_{\ell m}  \\
R^{(i)}_\ell(z^\alpha) &=& N_\ell^{\tau\tau (i)} \sum_\alpha F_\ell^{\alpha\beta} w^{(i)}_\ell(z^\beta)\label{eq:mean_response}
\ea
and a short calculation shows that the signal and noise power spectra are given by:
\ba
\langle \hE^{(i)*}_{\ell m} \hE^{(j)}_{\ell' m'} \rangle_{\rm noise} &=& N_\ell^{\tau\tau (i)} \delta_{ij} \delta_{\ell\ell'} \delta_{mm'}  \\
\langle \hE^{(i)*}_{\ell m} \hE^{(j)}_{\ell' m'} \rangle_{\rm signal} &=& \lambda^{(i)}_\ell N_\ell^{\tau\tau (i)} \delta_{ij} \delta_{\ell\ell'} \delta_{mm'}
\ea
Because estimators $i\ne j$ are uncorrelated, it is straightforward to calculate total signal-to-noise: we simply sum over eigenmodes,
\be
S/N = \left[\frac{\fsky}{2} \sum_{i\ell} (2\ell+1) (\lambda^{(i)}_\ell)^2\right]^{1/2}   \label{eq:sn2_eigenmode_sum} 
\ee
The eigenmode construction is convenient for studying signal-to-noise properties in the multifield case: one can compute the total $S/N$ achievable
in principle with quadratic estimator, or count the number of eigenmodes with significant $S/N$ in the sum in Eq.~(\ref{eq:sn2_eigenmode_sum}).

\section{Power spectrum calculations}
\label{app:power_spectrum_calc}

In this appendix, we will describe our procedure for calculating the power spectra $C_\ell^{X_0Y_1}$, where $X,Y\in\{T,E\}$.
We also give the explicit form of the $T_1$-field, which was left as a functional derivative in Eq.~(\ref{eq:sec3_fd}) and not evaluated explicitly.

We will work in the synchronous gauge where $h$ and $\eta$ denote the scalar degrees of freedom of the metric.
Overdots denote derivatives with respect to conformal time.
We denote the visibility function by $g(\chi) = \dot \tau e^{-\tau}$.
Further notation in this appendix follows \cite{Zaldarriaga:1996xe}: $v_b$ is the baryon velocity, 
$\Delta_{T,\ell}$ and $\Delta_{P,\ell}$ denote the Legendre expansion of the local temperature and polarization, respectively, $\Pi=\Delta_{T2}+\Delta_{P0}+\Delta_{P2}$ is the source for polarization generated by Thomson scattering,
and $\alpha=(\dot h+6\dot\eta)/(2k^2)$.

Let us begin by calculating $C_\ell^{E_0E_1}$.  The line-of-sight integral for polarization (Eq.~(\ref{eq:line_of_sight_qu})) can be rewritten in harmonic space
\cite{Zaldarriaga:1996xe} as:
\ba
a^{E}_{\ell m} &=& (4\pi i^\ell)
 \frac{3}{4} \sqrt{\frac{(\ell+2)!}{(\ell-2)!}} 
\int d^3\k\,
Y_{\ell m}^*(\hk)  \nn \\
&&
\hskip 0.2in \times \int d\chi\,  g(\chi) \, \frac{j_\ell(k\chi)}{(k\chi)^2} \Pi(\chi,\k)  \label{eq:app_los_e}
\ea
Because the perturbations are linear, $\Pi(\chi,\k)$ depends linearly on the initial curvature perturbation $\zeta(\k)$
and so we can rewrite Eq.~(\ref{eq:app_los_e}) in the form:
\be\label{eq:Elm_appb}
a^{E}_{\ell m} = 4\pi i^\ell \int d^3\k\, \Delta_\ell^E(k) \zeta(\k) Y_{\ell m}^*(\hk)
\ee
This representation encodes all the evolution in a single transfer function $\Delta^E_\ell$ and will be convenient
for computing power spectra below.

The response field $E_1$ can be treated analogously.
We rewrite the line-of-sight integral for $E_1$ (Eq.~(\ref{eq:line_of_sight_qu1})) in harmonic space:
\ba
a^{E_1(\chi)}_{\ell m} &=& (4\pi i^\ell)
 \frac{3}{4} \sqrt{\frac{(\ell+2)!}{(\ell-2)!}} 
\int d^3\k\,
Y_{\ell m}^*(\hk)  \nn \\
&&
\times \Bigg[
e^{-\tau(\chi)} \frac{j_\ell(k\chi)}{(k\chi)^2} \Pi(\chi,\k) \nn \\
&&
- \int_\chi^\infty d\chi'  g(\chi') \, \frac{j_\ell(k\chi')}{(k\chi')^2} \Pi(\chi',\k) \Bigg]
\ea
and define a transfer function $\Delta_\ell^{E_1(z)}$ so that:
\be
a^{E_1(z)}_{\ell m} = 4\pi i^\ell \int d^3\k\, \Delta_\ell^{E_1(z)}(k) \zeta(\k) Y_{\ell m}^*(\hk)
\ee
The transfer functions $\Delta_\ell^E(k), \Delta_\ell^{E_1(z)}(k)$ can be computed using CAMB, and
$C_\ell^{E_0E_1}$ is given by:
\be
C_\ell^{E_0E_1(z)} = (4\pi)^2 \int k^2dk\, \Delta_\ell^{E}(k) \Delta_\ell^{E_1(z)}(k) P_\zeta(k)
\ee
This concludes our calculation of $C_\ell^{E_0E_1}$; let us now discuss the temperature field.

The line-of-sight integral for temperature \cite{Zaldarriaga:1996xe} is given by
\ba\label{eq:Tlm_appb}
a_{\ell m}^{T} &=& (4\pi i^\ell) \int d^3\k\, Y_{\ell m}^*(\hk)
\int d\chi\, j_\ell(k\chi)  \nn \\
&& \times \Bigg( e^{-\tau(\chi)} Q_0(\chi,\k) + g(\chi) Q_1(\chi,\k) \nn \\
&& \hskip 0.25in + \dot g(\chi) Q_2(\chi,\k) + \ddot g(\chi) Q_3(\chi,\k) \Bigg)  \label{eq:app_los_t}
\ea
where
\ba
Q_0 &=& \dot\eta + \ddot\alpha \\
Q_1 &=& \Delta_{T,0} + 2\dot\alpha + \frac{\dot v_b}{k} + \frac{\Pi}{4} + \frac{3\ddot\Pi}{4k^2} \\
Q_2 &=& \alpha + \frac{v_b}{k} + \frac{3\dot\Pi}{4k^2} \\
Q_3 &=& \frac{3\Pi}{4k^2} 
\ea
(Note that the quantity in parentheses in Eq.~(\ref{eq:app_los_t}) is what we called $S_T$ previously in Eq.~(\ref{eq:s_t_def}).)

To get $T_1$ from $T$, we take the functional derivative (see Eq.~(\ref{eq:sec3_fd})):
\ba\label{eq:t1_horrible}
a^{T_1(z)}_{\ell m} &=& \int_{\chi(z)}^\infty d\chi\, \frac{\delta a^{T}_{\ell m}}{\delta \tau(\chi)}   \\
&=& (4\pi i^\ell) \int d^3\k\, Y_{\ell m}^*(\hk) \Bigg[ e^{-\tau(\chi(z))} B_0(\chi(z),\k)  \nn \\
 && + g(\chi(z)) B_1(\chi(z),\k) + \dot g(\chi(z)) B_2(\chi(z),\k) \nn \\
 && - \int_{\chi(z)}^\infty d\chi \Bigg( e^{-\tau(\chi)} Q_0(\chi,\k) + g(\chi) Q_1(\chi,\k) \nn \\
 && + \dot g(\chi) Q_2(\chi,\k) + \ddot g(\chi) Q_3(\chi,\k) \Bigg) j_\ell(k\chi) \Bigg] \nn
\ea
where:
\ba
B_0 &=& Q_1 j_\ell(k\chi) + \frac{\partial}{\partial\chi}(Q_2 j_\ell(k\chi)) + \frac{\partial^2}{\partial\chi^2}(Q_3 j_\ell(k\chi)) \nn \\
B_1 &=& Q_2 j_\ell(k\chi) + \frac{\partial}{\partial\chi}(Q_3 j_\ell(k\chi)) \nn \\
B_2 &=& Q_3 j_\ell(k\chi)
\ea
There is a subtlety here: when we take the functional derivative, we only differentiate the ``explicit'' dependence of $T(\n)$ on
$\tau(\chi)$ through the factors of $e^{-\tau}$, $g$, $\dot g$ and $\ddot g$ in Eq.~(\ref{eq:app_los_t}).  We treat the source terms $Q_i$
as independent of $\tau$, but this is actually an approximation since $\tau$ does appear in the evolution equation for these quantities.
This approximation is equivalent to dropping some double-scattering terms which are suppressed by one power of $\tau$; we checked
directly that the approximation is good to a few percent.  (The same approximation was made previously in polarization.)

To compute power spectra, we define transfer functions $\Delta_\ell^{T}$ and $\Delta_\ell^{T_1(z)}$ so that
\ba
a^{T}_{\ell m} &=& 4\pi i^\ell \int d^3\k\, \Delta_\ell^T(k) \zeta(\k) Y_{\ell m}^*(\hk) \\
a^{T_1(z)}_{\ell m} &=& 4\pi i^\ell \int d^3\k\, \Delta_\ell^{T_1(z)}(k) \zeta(\k) Y_{\ell m}^*(\hk)  \nn
\ea
As in the polarization case, we compute the transfer functions using CAMB.  The power spectra which contain temperature are then given by:
\ba
C_\ell^{T_0T_1(z)} &=& (4\pi)^2 \int k^2dk\, \Delta_\ell^{T}(k) \Delta_\ell^{T_1(z)}(k) P_\zeta(k)  \nn \\
C_\ell^{T_0E_1(z)} &=& (4\pi)^2 \int k^2dk\, \Delta_\ell^{T}(k) \Delta_\ell^{E_1(z)}(k) P_\zeta(k)  \nn \\
C_\ell^{E_0T_1(z)} &=& (4\pi)^2 \int k^2dk\, \Delta_\ell^{E}(k) \Delta_\ell^{T_1(z)}(k) P_\zeta(k) \nn
\ea
\vfill

\end{document}